\documentstyle[twocolumn,prl,aps,epsfig]{revtex}

\begin{document}
\draft

\def\be{\begin{equation}}
\def\ee{\end{equation}}
\def\bfi{\begin{figure}}
\def\efi{\end{figure}}
\def\bea{\begin{eqnarray}}
\def\eea{\end{eqnarray}}

\title{On the connection between off-equilibrium response and statics in non disordered coarsening systems}

\author{F.Corberi$^a$, E.Lippiello$^b$ 
\and M.Zannetti$^c$ 
}                     
%
%
\address{{\it Istituto Nazionale per la Fisica della Materia, 
Unit\`a di Salerno} and \\ 
{\it Dipartimento di Fisica, Universit\`a di Salerno,
84081 Baronissi (Salerno), Italy}\\
$^a$ corberi@na.infn.it\\
$^b$ lippiello@sa.infn.it\\
$^c$ zannetti@na.infn.it}

\maketitle

\begin{abstract}

The connection between the out of equilibrium linear response function
and static properties established by Franz, Mezard, Parisi and Peliti
for slowly relaxing systems is analyzed in the context of phase ordering
processes.
Separating the response in the bulk of domains from interface
response, we find that in order for the connection to hold 
the interface contribution must be asymptotically negligible.
How fast this happens depends on the competition between interface
curvature and the perturbing external field in driving domain growth.
This competition depends on space dimensionality and
there exists a critical value $d_c=3$ below which the interface
response becomes increasingly important eventually invalidating
the connection between statics and dynamics as the limit
$d=1$ is reached. 
This mechanism is analyzed numerically for the Ising model with $d$ 
ranging from $1$ to $4$ and analytically for a
continuous spin model with arbitrary dimensionality.
\end{abstract}

\section{Introduction} \label{intro}

The off-equilibrium character of the time evolution of a system undergoing
a phase ordering process, such as a ferromagnet quenched below the 
critical point, is clearly manifested by the aging property observed in
the response function. If the system is cooled in zero field and
left in the low temperature phase for a time $t_w$ before applying an
external field, for $t_w$ sufficiently large the time dependent magnetization 
displays a behavior of the type
\be 
M(t,t_w) \simeq M_{st}(t-t_w)+M_{ag}(t,t_w)
\label{1.1}
\ee
where $M_{st}(t-t_w)$ is a stationary time translation invariant (TTI)
contribution and the remaining term $M_{ag}(t,t_w)$ is the aging 
contribution obeying the scaling form
\be
M_{ag}(t,t_w) = t_{w}^{-a} {\cal M}\left ( \frac{t}{t_w} \right) .
\label{1.10}
\ee
A structure of the same type shows up also in the autocorrelation
function given by
\be 
C(t,t_w) \simeq C_{st}(t-t_w)+C_{ag}\left ( \frac{t}{t_w}\right) .
\label{1.2}
\ee
Behaviors like~(\ref{1.1}) and~(\ref{1.2}) are common features
of slow relaxation and are the object of very intensive study especially in
glassy systems, with and without disorder~\cite{Bouchaud97}. 

In the case of systems evolving
via domain coarsening, structures of this type
can be readily interpreted in terms of two independent variables responsible,
respectively, of the fast thermal fluctuations within domains and of the
slow out of equilibrium interface dynamics. The 
splitting of the order parameter into thermal and ordering 
components was used some time ago~\cite{MVZ} as the key ingredient
in the theory of phase ordering.
Therefore, the stationary contributions in~(\ref{1.1}) and~(\ref{1.2}) 
are due to equilibrium thermal fluctuations in the bulk of domains,
while the aging terms come from 
the remaining out of equilibrium fluctuations occurring at the
passage of interfaces~\cite{Berthier99,Franz00}.

In the study of glassy systems, along with the realization that in 
these systems the out of equilibrium properties are of
foremost importance, recently there has been a pair of developments which have
further enhanced the interest in the dynamics of slow relaxation. The first
has been the off-equilibrium generalization of the fluctuation dissipation
theorem (FDT), first derived by Cugliandolo and Kurchan~\cite{Cugliandolo93} 
in the context of
mean field models for spin glasses. This amounts to the statement that for
$t_w \rightarrow \infty$ the magnetization depends on the time
variables only through the autocorrelation function
\be 
M(t,t_w) = M[C(t,t_w)]
\label{1.3}
\ee
and the deviation from the ordinary FDT can be expressed through the so
called fluctuation dissipation ratio (FDR)
\be
X(C) = -\frac{dM(C)}{dC}
\label{1.4}
\ee
which obeys $X(C)=1$ in equilibrium. The second is a theorem by 
Franz, Mezard, Parisi and Peliti (FMPP)~\cite{FMPP} which allows
to retrieve the 
structure of the equilibrium state from dynamic properties during
relaxation. Under certain hypothesis, they have established the identity
\be
\left . \frac{dX(C)}{dC} \right ] _{C=q} = P(q)
\label{1.5}
\ee
where $P(q)$ is the overlap probability distribution in the equilibrium 
state~\cite{Mezard87}.
This latter development is of particular significance, since it opens a 
way around the difficulty of static computations for systems with complex 
equilibrium states.

In this context, the phase ordering process in pure systems is of 
considerable interest as a simplified framework where the chain of connections
{\it aging-FDR-statics} can be analyzed and tested. 
The main point is that the phenomenology of phase ordering displays the
typical features~(\ref{1.1}) and~(\ref{1.2}) of slow relaxation, and that the
structure of the equilibrium state is exactly known, thus allowing for
a detailed investigation of the relation between statics and dynamics. 
Work in this direction~\cite{Berthier99,Barrat98,Parisi99} 
has led to
the conclusion that the aging term in the response function
does not play any role asymptotically, therefore relegating the connection
between static and dynamic properties in the somewhat trivial bulk
contribution. The argument is based on the statement that interface
response comes only from the spins on the border of growing domains, yielding
the upper bound
\be
M_{ag}(t,t_w) \leq \rho_I(t_w) \simeq L^{-1}(t_w)
\label{1.6}
\ee
where $\rho_I(t_w)$ is the interface density and $L(t_w) \sim t_w^{1/z}$ 
is the typical domain size. This fits into the 
form~(\ref{1.10}) with $a=1/z$, where $z$ is the growth exponent.

\begin{figure}
\resizebox{0.5\textwidth}{!}{%
  \includegraphics{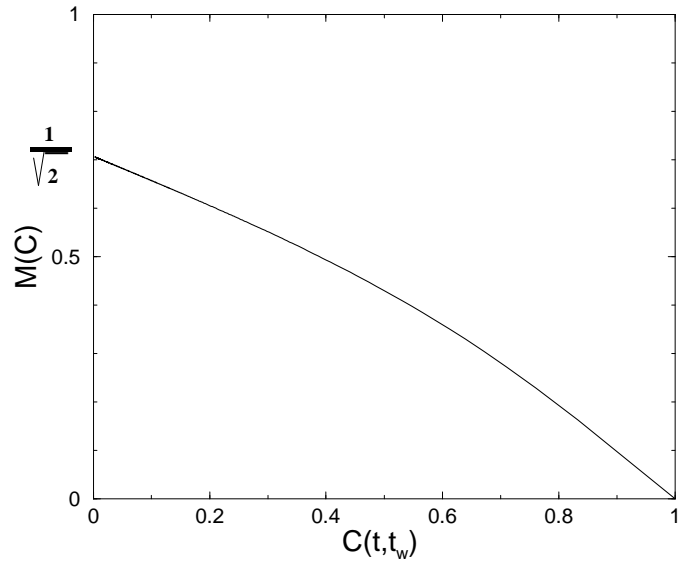}
}
\caption{$M(C)$ for the $d=1$ Ising model
with $J=\infty $.}
\label{isinguno}
\end{figure}

However, this picture is at variance with 
exact analytical results for the one dimensional Ising 
model~\cite{Godreche00,Lippiello00} in the limit of infinite ferromagnetic
coupling (the reason for taking this limit rather than the zero temperature
limit will be discussed in Section 6). 
In this 
case one finds the opposite situation, namely there is no 
bulk response, while the interface response obeys~(\ref{1.10}) with
$a=0$ and 
\be
{\cal M}\left (\frac{t}{t_w}\right ) = \frac{\sqrt{2}}{\pi} \arctan 
\sqrt{\frac{t}{t_w} -1}
\label{1.20}
\ee
yielding a finite asymptotic value independent of $t_w$
\be
\lim_{t \rightarrow \infty} M_{ag}(t,t_w) = \frac {1}{\sqrt{2}}.
\label{1.21}
\ee
Similarly, there is no stationary term in the autocorrelation function,
while the aging term is given by~\cite{Bray89,Prados97}
\be
C_{ag}\left (\frac{t}{t_w}\right ) = \frac{2}{\pi} \arcsin 
\left (\frac{2}{1+\frac{t}{t_w}} \right ).
\label{1.22}
\ee
Hence, eliminating $t/t_w$ between~(\ref{1.20}) and~(\ref{1.22}) one
finds (Fig.~\ref{isinguno})
\be
M(C)= \frac{\sqrt{2}}{\pi} \arctan \left [\sqrt{2} \cot \left (\frac{\pi}{2}C
\right ) \right ]
\label{1.23}
\ee
showing that the response function obeys~(\ref{1.3}) for
any $t_w$ giving rise to a non trivial FDR which, however, 
leads to a violation of
the connection~(\ref{1.5}) between static and dynamic properties. This
result indicates that the aging part
of the response function for coarsening systems might not be as 
simple as~(\ref{1.6}). In order to
address this problem~\cite{Corberi01}, we have analyzed the behavior
of the interface contribution to the response function as dimensionality is
varied, finding through simulations for Ising spins and a phenomenological
model for continuous spins that the interface contribution is indeed less
trivial than hitherto believed. We find that the scaling form~(\ref{1.10})
holds with a scaling function and an exponent $a$ which depend on 
dimensionality, providing a unified
coherent picture for the diverse behaviors
observed at different dimensionalities. More specifically, we find 
that $d_c=3$ is the critical dimensionality such that:

\noindent {\it i)} for $d> d_c$ the response is actually due only to the
polarization of the spins at the interfaces making~(\ref{1.6}) to hold

\noindent {\it ii)} for $d< d_c$ there is a new and
non trivial behavior of the response function due to the 
competition in the motion of interfaces between the drive of the curvature,
aiming to minimize surface tension, and the drive of the external field,
aiming to minimize the magnetic energy of domains.

The paper is organized as follows. In Section 2 general concepts
about the structure of phase space and time evolution 
are reviewed. Section 3 and Section 4 are devoted, respectively,
to the relaxation process dominated by the fast degrees of freedom 
leading to equilibration and to the phase ordering process which is,
conversely, dominated by the slow out of equilibrium degrees of
freedom. Section 5 contains a short account of  the FMPP scheme for
the connection between static and dynamic properties. Section 6 and
Section 7 contain results, respectively, for the Ising model
in $d=1$ and in higher dimensions. The model for continuous spins is
presented in Section 8 and concluding remarks are made in Section 9.

\section{Structure of phase space} \label{component}

Let us consider a spin system with hamiltonian ${\cal H}[s_i]$,
for instance the ferromagnetic Ising model, in contact with a 
thermal reservoir 
at the temperature $T$.
Below the critical temperature $T_c$ configuration space
breaks up into ergodic components~\cite{Palmer82}
\be
\Omega =\left (U_\alpha \Omega _\alpha \right )U\Omega _b
\label{2.7}
\ee
where by $\Omega _\alpha $, with $\alpha =\pm$ we have denoted
the basins of attraction of pure states and by $\Omega _b$ the
boundary between them~\cite{Kurchan96}. The Gibbs state
\be
\rho_{G}(\omega)=\frac{1}{Z} e^{-\frac {1}{T} {\cal H}(\omega)}
\label{2.1}
\ee
is the mixture of the two broken symmetry pure states
\be
\rho_{G}(\omega)=w_+\rho_+(\omega)+w_-\rho_-(\omega) 
\label{2.2}
\ee
where
$Z=\sum _{\omega \in \Omega} e^{-\frac {1}{T} {\cal H}(\omega)}$,
$\omega =[s_i]$ is a spin configuration, the pure states are given by
\be
\rho_\alpha (\omega) = \left \{ \begin{array}{ll}
                   \frac{1}{Z_\alpha} e^{-\frac{1}{T} {\cal H}(\omega)}
                              & \mbox{, if $\omega\in \Omega _\alpha$} \\
                   0          & \mbox{, if $\omega\notin \Omega _\alpha$}
                   \end{array}
              \right .
\label{2.11}
\ee
and $w_+=w_-=1/2$.
In each pure state there is spontaneous magnetization
\be
m_\pm =\frac {1}{N} \sum _i \langle s_i \rangle_\pm =\pm m_T
\label{2.3}
\ee
and a finite correlation length $\xi _T$ which does not depend on the sign
of the state and is related to the relaxation time within the pure
state by
\be
\tau ^{1/z}\sim \xi _T.
\label{2.500}
\ee

In what follows $\tau $
will characterize the time scale of fast relaxation.

Alternatively, 
defining the overlap of two configurations by
\be
Q(\omega,\omega ')=\frac{1}{N}\sum _i s_i s_i'
\label{2.4}
\ee
the structure of a state $\rho (\omega )$ can be characterized through 
the probability~\cite{Mezard87}
that $Q(\omega,\omega ')$ takes the value $q$ when $\omega$ and
$\omega '$ are configurations of two independent copies of the system
\be
P(q)=\sum _{\omega,\omega '}\rho(\omega)\rho(\omega ')
\delta \left (Q(\omega,\omega ')-q\right ).
\label{2.5}
\ee
Using (\ref{2.2}) and (\ref{2.3}), the overlap probability function in the
Gibbs state is given by
\be
P_{G}(q)=(w_+^2+w_-^2)\delta (q-m_T^2)+2w_+w_-\delta (q+m_T^2)
\label{2.6}
\ee
where the mixed character of the state is revealed by the presence of the
second $\delta$-function in the right hand side. 
For future reference, notice that from~(\ref{2.5}) follows
\be
\int dq P(q) q = \frac{1}{N} \sum_i \left < s_{i} \right >^2.
\label{2.61}
\ee

Let us now consider the instantaneous quench process, where the system is
initially prepared in some initial state $\rho_0(\omega)$ and, 
at the time $t=0$,
is put in contact with the thermal reservoir at the temperature $T<T_c$. 
Taking $t>\tau$, the measure over $\Omega $ is given by
\be
\rho(\omega ,t)=\sum _\alpha \rho_0 (\Omega _\alpha )
\rho_\alpha (\omega )
+\rho_b(\omega ,t)
\label{2.18}
\ee
where $\rho_0 (\Omega _\alpha )=\sum _{\omega \in \Omega _\alpha }
\rho_0(\omega)$ and $\rho _b (\omega ,t)$ is the measure over the
boundary.
Similarly, for the joint probability at times $t>t'>\tau $ we may write
\be
\rho(\omega 't',\omega t)=\sum _\alpha \rho_0(\Omega _\alpha )
\rho_\alpha (\omega ',\omega,t-t')+\rho_b(\omega ' t',\omega t)
\label{2.19}
\ee
where $\rho_\alpha (\omega ', \omega,t-t')$ is the 
TTI pure state joint probability. From~(\ref{2.18}) and~(\ref{2.19})
it is quite clear that the properties of the system following a quench
below $T_c$ are sensitive~\cite{Palmer82} to the choice of the initial condition
$\rho_0(\omega )$, specifically to the weight given at the time $t=0$ 
to the different components. 

At the level of the observables of interest, like magnetization
$m(t)=\langle s_i (t)\rangle$ and correlation function
$C(\mid i-j\mid,t,t')=\langle s_i(t)s_j(t')\rangle
-\langle s_i(t)\rangle \langle s_j(t')\rangle$,
where space translation invariance is assumed to hold, the above results
translate in the following way. From~(\ref{2.18}) follows that for $t>\tau $
\be
m(t)=\sum _\alpha \rho_0(\Omega _\alpha)  m_\alpha +m_b (t)
\label{2.22}
\ee
where $m_\alpha $ is the
equilibrium value of the magnetization in the pure states given by~(\ref{2.3}).
Next, assuming that on the boundary $m_b(t)=0$, for $t>t'>\tau $ and
from~(\ref{2.19}) we have
\bea
C(\mid i-j\mid,t,t')&=&\left [\sum_{\alpha}\rho_0(\Omega _\alpha)\right ] 
            C_{ps}(\mid i-j\mid,t-t') \nonumber \\
                    &+& C_b(\mid i-j\mid,t,t') + \Delta m
\label{2.23}
\eea
where
\be
C_{ps}(\mid i-j\mid,t-t')=\langle s_i(t)s_j(t')\rangle_\alpha -m_\alpha ^2
\label{2.24}
\ee
is the TTI correlation function in the equilibrium pure states which,
for pure states related by symmetry,
is independent of $\alpha $. Furthermore
\be
C_b(\mid i-j\mid,t,t')=\langle s_i(t)s_j(t')\rangle _b
\label{2.25}
\ee
is the correlation function on the boundary and
\be
\Delta m=\sum _\alpha \rho_0(\Omega _\alpha) m_\alpha ^2
-\left [ \sum _\alpha \rho_0(\Omega _\alpha) m_\alpha \right ]^2
\label{2.26}
\ee
gives  the fluctuation of the magnetization over pure states. Properties of
the pure state correlation function which will be needed in the following are
\be
C_{ps}(\mid i-j\mid=0,t-t'=0)=1-m_T^2
\label{2.27}
\ee
where we have used $s_i ^2=1$, and
\be
C_{ps}(\mid i-j\mid,t-t')=0
\label{2.28}
\ee
for $\mid i-j\mid>\xi _T$ or $t-t'>\tau$.

\section{Fast process: Relaxation to equilibrium} \label{fastrelax}

\begin{figure}
\resizebox{0.5\textwidth}{!}{%
  \includegraphics{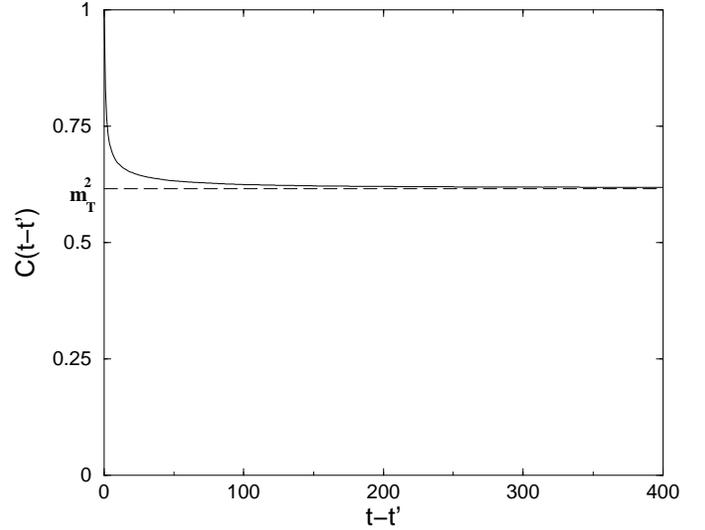}
}
\caption{Autocorrelation
function for the $d=2$ Ising model with $J=1$ quenched to $T=0.969T_c$ 
($m_T^2=0.61584$) from the initial condition 
$\rho_0(\omega)= \frac{1}{2} \delta(\omega- \omega_{+})
+ \frac{1}{2} \delta(\omega- \omega_{-})$ with 
$\omega_{+}=[s_i=1, \forall i]$ and $\omega_{-}=[s_i=-1, \forall i]$.}
\label{c_purestate}
\end{figure}

Let us now adjust the initial condition of the quench in order to have
relaxation to the Gibbs
state~(\ref{2.1}). From~(\ref{2.2})  and~(\ref{2.11}) the Gibbs state is the
mixture of pure states
with weights $ w _\alpha = Z_\alpha / Z $.
Hence, according to~(\ref{2.18}) relaxation to the Gibbs state can take place
only if the initial
condition is such that
\begin{equation}
\left \{ \begin{array}{ll}
                   \rho_0(\Omega _\alpha )=\frac{Z _\alpha }{Z} \nonumber \\
                   \rho_0(\Omega _b )=0.
                   \end{array}
              \right .
\label{3.1}
\end{equation}
With such an arrangement, equilibrium is reached in the time scale $
\tau $. Having now $ \sum
_\alpha \rho_0(\Omega _\alpha)   =1 $, with  $\rho_0(\Omega _\alpha) =1/2 $  
independent of $\alpha$,
~(\ref{2.22}) and
~(\ref{2.23}) simplify to $ m(t) = 0 $ and
\begin{equation}
C( \vert i-j \vert,t-t') = C _{ps} ( \vert i-j\vert, t-t') + m _T ^2
\label{3.2}
\end{equation}
implying the no clustering property $ C(\vert i-j\vert, t-t') \geq m_T ^2 $.
This behavior is illustrated in Fig.~\ref{c_purestate} 
depicting the autocorrelation
function $C(t-t')=C(\vert i-j \vert=0,t-t')$ 
for the $d=2$ Ising model quenched to $T=0.969 T_c$ 
with the
initial condition $\rho_0(\omega)= \frac{1}{2} \delta(\omega- \omega_{+})
+ \frac{1}{2} \delta(\omega- \omega_{-})$ concentrating the measure on
the bottom of the basins of attraction, where 
$\omega_{+}=[s_i=1, \forall i]$ and $\omega_{-}=[s_i=-1, \forall i]$.

An observation important for what follows is that the form~(\ref{3.2}) 
of the correlation function corresponds to the
splitting of the spin variable into the sum of two statistically independent
components
\begin{equation}
s _i(t) = \psi _ i(t) + \sigma
\label{3.3}
\end{equation}
where $ \psi _i(t) $ obeys the statistics of equilibrium thermal
fluctuations in the pure state with expectations
\begin{equation}
\left \{ \begin{array}{ll}
                   \langle \psi _i(t) \rangle_\alpha =0 \nonumber \\
 \langle \psi _i(t) \psi _j(t')\rangle_\alpha = C _{ps}( \vert i-j\vert, t-t').
                   \end{array}
              \right .
\label{3.4}
\end{equation}
Instead, $\sigma$ is a time independent random variable, the 
ordering component, which takes the values $ m _\alpha $ with  
probabilities $ p(m _\alpha ) = \rho_0(\Omega _\alpha) $ determined by the
initial condition. Denoting the latter average by an overbar we have
\begin{equation}
\left \{ \begin{array}{ll}
                 \overline \sigma=0   \nonumber \\
                 \overline{\sigma ^2} = m_T^2
                   \end{array}
              \right .
\label{3.5}
\end{equation}
and~(\ref{3.2}) can be rewritten as 
\begin{equation}
C( \vert i-j\vert,t-t') = \langle \psi _i(t) \psi _j(t') \rangle _\alpha + 
\overline{\sigma ^2}. 
\label{3.6}
\end{equation}

The next step is to study the response of the system to a perturbation. 
Suppose that at the
time $ t _w $ after the quench the hamiltonian ${\cal H}(\omega )$
is changed into 
\be
{\cal H}_h(\omega)={\cal H}(\omega) -{\cal H}_1(\omega)
\label{2.12}
\ee
where
\be
{\cal H}_1(\omega)=\sum _i h_is_i
\label{2.13}
\ee
is an uncorrelated
gaussian random field (RF) with expectations
\begin{equation}
\left \{ \begin{array}{ll}
               E_h(h_i) = 0 \nonumber \\
               E_h(h_ih_j)=h_0^2\delta_{ij}.
               \end{array}
            \right .
\label{2.17}
\end{equation}
In the RF Ising model the lower critical dimensionality is raised from
$d_L=1$ to $d_L=2$. Hence, for $d>2$ the component structure~(\ref{2.7})
of configuration space is not modified by the presence of the RF.
Due to the presence of the
perturbation the system will relax to a new equilibrium state
\begin{equation}
\rho ^*(\omega) = \sum _\alpha \rho_0(\Omega _\alpha) 
\rho _{\alpha ,h} (\omega)
\label{3.7}
\end{equation}
which is neither the perturbed 
\be
\rho_{G,h}(\omega)=\frac{1}{Z_h}e^{-\frac{1}{T}{\cal H}_h(\omega )}
\label{2.14}
\ee
nor the unperturbed Gibbs state~(\ref{2.1}), since the
perturbation is present in the pure states, but not in the weights 
$\rho_0(\Omega _\alpha) $. 
We will be
interested in the linear response to the perturbation, since already in
the simple context of fast relaxation it is possible to identify 
some of the basic
elements of the connection between static and dynamic properties to be
discussed in Section 5. Let us then consider the staggered
magnetization defined by
\begin{equation}
M(t-t _w)= \frac{T}{N h _0 ^2} 
E_h \left [ \langle {\cal H} _1(t) \rangle _h \right ]=
\frac{T}{N h _0 ^2} E_h \left [ \sum _i \langle s _i(t) \rangle _h h _i 
\right ]
\label{3.8}
\end{equation}
where $ \langle \cdot \rangle _h $ denotes 
the thermal
average for a given RF realization.
Taking $ t _w> \tau $, i.e. switching on the perturbation after the
unperturbed system has reached equilibrium, this quantity depends only on 
the time difference. 
Since by definition $ \langle \psi_i(t) \rangle =0$,  
 we may write $ \langle s _i(t) \rangle _h =
\overline{\sigma _i (t)} ^h $ where, due to the RF, the ordering component $
\sigma $ acquires a site and time dependence. Expanding up to linear 
order in the field and recalling that the unperturbed average 
$\overline{\sigma}$ vanishes we have
\begin{equation}
\overline{\sigma _i(t)} ^h =  \sum _j
\chi (\vert i-j \vert,t- t _w) h _j + {\cal O}(h ^2)
\label{3.9}
\end{equation}
where from~(\ref{3.7}) 
\be
\chi (\vert i-j \vert,t- t _w) = \sum_{\alpha} \rho_0(\Omega _\alpha) 
\chi_{\alpha} (\vert i-j \vert,t- t _w)
\label{3.90}
\ee
and for pure states related by symmetry 
$\chi_{\alpha} (\vert i-j \vert,t- t _w)$ 
is independent of $\alpha$. Inserting~(\ref{3.9}) and~(\ref{3.90}) 
into~(\ref{3.8}) we obtain 
\begin{equation}
M(t-t _w) = T \chi_{\alpha} (t-t _w)
\label{3.10}
\end{equation}
where $\chi_{\alpha} (t-t _w) = \chi_{\alpha} (\vert i-j \vert =0,t- t _w)$.
Hence,
$\lim _{t \to \infty} M (t-t _w) = M_{\alpha} = T \chi_{\alpha}$
where, using~(\ref{2.27}), the static susceptibility in either one of the
pure states is given by 
\begin{equation}
\chi_{\alpha} = \frac{1}{T} \left [ 1- m _T ^2 \right ].
\label{3.14}
\end{equation}
This can also be rewritten as
\begin{equation}
M_{\alpha} = 1 - \int dq P_\alpha (q) q
\label{3.15}
\end{equation}
where
\begin{equation}
P_\alpha (q) = \delta (q-m _T ^2)
\label{3.16}
\end{equation}
is the overlap probability function of the unperturbed pure states. We call the
attention here on the point that this result is different from
what one obtains computing the susceptibility from the 
perturbed Gibbs state~(\ref{2.14}), which differs from~(\ref{3.7}) 
because the RF dependence
enters also in the weights $ w _{\alpha,h} $. In that case, in 
place of~(\ref{3.14}) one finds
\begin{equation}
\chi _{G} = \frac{1}{T} \left [ 1 - \langle s _i \rangle _{G} ^2 \right ]
\label{3.17}
\end{equation}
where $ \langle \cdot \rangle _ {G} $ denotes the average 
over the unperturbed
Gibbs state. Recalling~(\ref{2.61}), this can be rewritten as
\begin{equation}
\chi _{G} = \frac{1}{T} \left [1- \int dq P_G(q) q \right ]
\label{3.18}
\end{equation}
where now $ P_G(q) $ is the overlap probability function~(\ref{2.6}) of the Gibbs
state. However, the form ~(\ref{3.17}) or ~(\ref{3.18}) of the susceptibility
cannot be reached dynamically. That is, by switching on the perturbation at the
time $ t _w $ after the quench, the $ t \to \infty $ limit of the 
staggered magnetization is given by ~(\ref{3.14}) and not by\ ~(\ref{3.17}),
which gives $\chi _{G}=1/T$ since $\langle s _i \rangle _{G} =0$.

\begin{figure}
\resizebox{0.5\textwidth}{!}{%
  \includegraphics{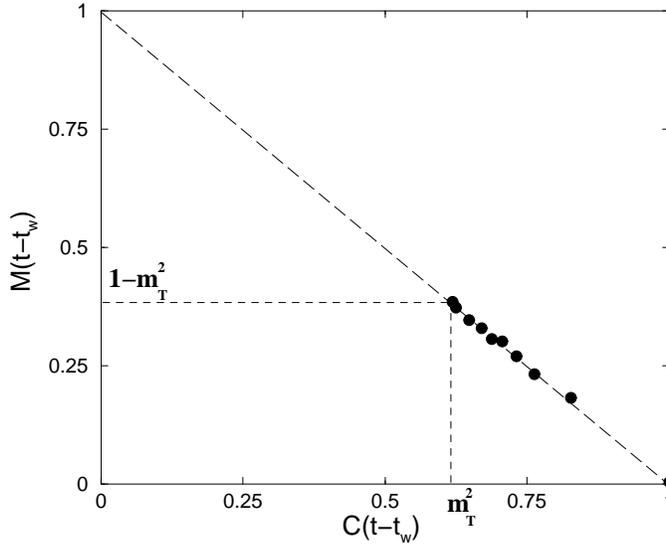}
}
\caption{$M (t-t_w)$ against $C(t-t_w)$ for the same quench as 
in Fig.~\ref{c_purestate}.
The horizontal dashed line represents the continuation of $M(C)$
into the unphysical region $C<m_T^2$.}
\label{comm_wis}
\end{figure}

Next, let us turn to the dynamical side of~(\ref{3.10}) and let us show,
for pedagogical purposes, that~(\ref{3.14}) and~(\ref{3.15}) can also be
obtained from a dynamical object like the linear response function
\be
R(t-t') = \left. \frac{\delta \langle s_{i}(t) \rangle}
{\delta h_{i}(t')}   \right ]_{h=0}
\label{3.111}
\ee
without resorting to knowledge of the equilibrium state. The response
function $\chi_{\alpha} (t-t _w)$ entering~(\ref{3.10}) and 
$R(t-t')$ are related by
\begin{equation}
\chi_{\alpha} (t-t _w) = \int _{t _w} ^t dt' R(t-t') 
\label{3.11}
\end{equation}
and, given that the unperturbed system is in equilibrium,
the linear response function and the pure state autocorrelation function
are related by the FDT 
\begin{equation}
R(t-t')= \frac{1}{T} \frac{ \partial C_{ps} (t-t')}{\partial t'}.
\label{3.12}
\end{equation}
Since the constant term in~(\ref{3.2}) makes no contribution to the time
derivative, we may replace $ C_{ps} (t-t')$ by the full autocorrelation
function $ C (t-t')$, and inserting~(\ref{3.12}) in~(\ref{3.11}) we
find
\be
M (t-t_w)= \int _{C(t-t_w)}^1 dq = \left [ 1-C(t-t_w)\right ].
\label{3.19}
\ee
This shows that when FDT holds the time dependence of $M$, or $\chi$,
is entirely absorbed in the linear dependence on the autocorrelation function
(Fig.~\ref{comm_wis}). 

From~(\ref{3.19}) $M (t-t_w)$ reaches the equilibrium 
value~(\ref{3.14}) as $C(t-t_w)$ reaches the lower bound $m_T^2$.
Even though $C(t-t_w)$  cannot fall below this value, we may extend the
dependence of $M$ on $C$ into the unphysical region $C<m_T^2$ 
(horizontal dashed line in Fig.~\ref{comm_wis}) by rewriting~(\ref{3.19}) as
\be
M(C)= \int _C ^1 \theta (q-m_T^2)dq
\label{3.20}
\ee
where $\theta$ denotes the Heaviside step function. Integrating
by parts we find
\be
M(C) = [1- C\theta(C-m_T^2)] - \int _C ^1 dq \delta (q-m_T^2)q
\label{3.22}
\ee
which yields
\be
M(C) = \left \{ \begin{array}{ll}
                   1-C
                              & \mbox{, for $m_T^2< C \leq 1$} \\
                   M_{\alpha}=1-m_T^2          & \mbox{, for $C \leq m_T^2$}.
                   \end{array}
              \right .
\label{3.21}
\ee
Taking $C=0$, from~(\ref{3.22}) we find
$M_{\alpha} = 1 - \int _0 ^1 dq \delta (q-m_T^2)q$
recovering~(\ref{3.15}). In order to understand this result, notice
that~(\ref{3.20}) can be regarded 
as obtained from~(\ref{3.11}) with the FDT in the modified form
\begin{equation}
R(t-t')= \frac{X[C(t-t')]}{T} \frac{ \partial C (t-t')}{\partial t'}
\label{3.222}
\end{equation}
where $X(C)=\theta(C-m_T^2)$ is the FDR introduced in~(\ref{1.4}). 
Rewriting~(\ref{3.22}) as
\be
M(C) = [1- C X(C)] - \int _C ^1 dq \frac{dX(q)}{dq} q
\label{3.223}
\ee
and taking again $C=0$ we find 
$M_{\alpha} = 1 - \int _0 ^1 dq \frac{dX(q)}{dq} q$.
Comparing with~(\ref{3.15}) then we find the relation~(\ref{1.5}) in the form
\be
\left. \frac{dX(C)}{dC} \right]_{C=q} = P_{\alpha}(q)
\label{3.225}
\ee
showing that the piece of static information contained in
$P_{\alpha}(q)$ is encoded into the relaxation properties
through the FDR. Although this may seem an artificial exercise, it will 
turn out useful in the understanding of the connection between static
and dynamic properties in the less trivial context of slow relaxation.

\section{Slow relaxation: phase ordering} \label{slow}

In the previous Section we have analyzed a quench process which yields
 equilibration in the Gibbs state within the time scale $ \tau $.
In order to achieve this the initial condition had to be chosen
according to~(\ref{3.1}). Now we turn to phase ordering~\cite{Bray94}, where
equilibrium is not reached within any finite time scale.
We will find out that in order to have a phase ordering process the
initial condition, in a sense, must be opposite to~(\ref{3.1})
with
\begin{equation}
\left \{ \begin{array}{ll}
                   \rho_0(\Omega _\alpha )=0 \nonumber \\
                   \rho_0(\Omega _b )=1.
                   \end{array}
              \right .
\label{4.1}
\end{equation}
Nonetheless, the fast equilibration process of the previous Section
will turn out to dominate the short time behavior of phase
ordering. 

In order to assess how the phase ordering
process fits in the general scheme of Section~\ref{component}, we 
relay on the behavior of the correlation function. Typically,
the initial state is taken as the infinite temperature equilibrium state
\begin{equation}
\rho_0 (\omega ) = \prod _i p(s_i)
\label{4.2}
\end{equation}
with $ p(s_i)= 1/2 $ which yields the uniform measure $ \rho_0(\omega) =
2 ^{-N} $. Taking the shortest time after the quench
$t'$ sufficiently larger than $ \tau $, the observed behavior
of the correlation function is well represented by the sum of two
contributions
\begin{equation}
C(\vert i-j \vert ,t,t') = C_{ps}(\vert i-j \vert ,t-t') +
C_{ag}(\vert i-j \vert ,t,t')
\label{4.3}
\end{equation}
where the first one is TTI and coincides with~(\ref{2.24}), 
while the second one displays aging through the scaling
behavior~\cite{Bray94,Furukawa89}
\begin{equation}
C_{ag}(\vert i-j \vert ,t,t') = m _T^2 F_{ag} \left (\frac{\vert i-j \vert}
{L(t')}, \frac{L(t)}{L(t')} \right).
\label{4.4}
\end{equation}
The characteristic length $ L(t) $ grows with the power law $ L(t) \sim
t^{1/z} $ where $ z=2 $ for non conserved order parameter, 
as it will be considered
in this paper. The scaling function $ F_{ag}(x,y) $ has the properties
\begin{equation}
\left  \{  \begin{array}{ll}
           F _{ag}(0,1)=1 \nonumber \\
           F _{ag}(x,1) \sim e^{-x^2} \quad , \quad \mbox{for $x \gg 1$} \nonumber \\
           F _{ag}(0,y) \sim   y^ {- \lambda} \quad \mbox{, for $y \gg 1$}
           \end{array}
       \right .
\label{4.8}
\end{equation}
with $\lambda > 0$.
This shows that, contrary to the previous case, now $ C( \vert i-j \vert,
t,t') $ becomes smaller than $ m_T ^2 $ and eventually vanishes
when $\vert i-j \vert$ or $t-t'$ becomes large. 
This, in turn, implies that with
the uniform initial state~(\ref{4.2}) condition~(\ref{4.1}) is realized,
otherwise from~(\ref{2.23}) follows that 
it would not be possible for $ C( \vert i-j \vert,t,t') $ 
to vanish at large distances or large time separations. 
Therefore we must have
\begin{equation}
C(\vert i-j \vert,t,t') = C _b(\vert i-j \vert,t,t')
\label{4.9}
\end{equation}
revealing that the structure~(\ref{4.3}) reflects a property of 
the evolution over the boundary $ \Omega _b $.
In the simplest case of the ferromagnetic system with two pure states, 
as we are considering, it is well known that the time
evolution of configurations is given by the coarsening of
domains of the two opposite equilibrium phases.
Taking $ t ' \gg \tau $ in order to separate
the time scales of fast and slow dynamics, within each domain 
the system is in equilibrium in either one of the two
pure states $ \rho_ \pm (\omega) $. Putting this together with~(\ref{4.3}),
since in the time regime $ t -t' \ll t' $ and for short distance
$ C_{ag} \simeq m_T ^2 $, we have that over short time and short
distances the correlation function is indistinguishable from~(\ref{3.2}).
Namely, we may regard the phase ordering process as a fast relaxation
process of the type considered in the previous Section, followed by a
quite different and much more slow relaxation taking place on the time
scale set by $ t' $. This suggests the separation of the spin variable
into a fast and a slow component, by generalization of the split~(\ref{3.3}) 
in the form~\cite{MVZ,Franz00}
\begin{equation}
s _i(t)= \psi _i(t) + \sigma _i(t)
\label{4.10}
\end{equation}
where, again, $ \psi _i $ and $ \sigma  _i $ are two statistically independent
variables. By analogy with ~(\ref{3.3}), we define the slow ordering component
by
\begin{equation}
\sigma _i(t) = \pm m_T
\label{4.11}
\end{equation}
according to the sign of the domain the site $ i $ belongs to at the time
$ t $. Then, the statistics of the one time properties of $\sigma _i(t)$ 
is determined by the relative occurrence of domains of either sign, which  
yields the time independent probability $p(\sigma_i)= 1/2$ 
and the expectations
\begin{equation}
\left \{ \begin{array}{ll}
                   \overline{\sigma _i(t)} = 0 \nonumber \\
                   \overline{\sigma_{i}^{2}(t)} = m _T^2
                   \end{array}
              \right .
\label{4.12}
\end{equation}
as in ~(\ref{3.5}).
However, since $ \sigma _i(t) $ changes sign whenever an interface passes
through the site $ i $, the two times statistics is determined by the out
equilibrium interface motion.
With this choice for the ordering component, $ \psi _i $  represents
the fast thermal fluctuations in the bulk of domains with the statistics
of the equilibrium pure states, which is independent of the sign of domains.
From ~(\ref{4.10}) then follows
\begin{equation}
C(\vert i-j \vert, t,t_w) = \langle \psi _i(t) \psi _j(t_w) \rangle _\alpha +
\overline{\sigma _i(t) \sigma _j(t_w)}
\label{4.14}
\end{equation}
and comparing with ~(\ref{4.3}) we can make the identifications
\begin{equation}
 \langle \psi _i(t) \psi _j(t_w) \rangle _\alpha =
C_{ps}(\vert i-j \vert, t-t_w)
\label{4.15}
\end{equation}
\begin{equation}
\overline{\sigma _i(t) \sigma _j(t_w)} = C_{ag}(\vert i-j \vert, t,t_w) .
\label{4.16}
\end{equation}

After surveying the unperturbed phase ordering process, let us go over
to the behavior of the staggered magnetization~(\ref{3.8}) after
switching on the perturbation~(\ref{2.13})  at the time $ t_w $.
As mentioned above, with the RF the lower critical dimensionality
is $d_L=2$.
We will assume that the external field is so small that the bound on the
size of domains imposed by the Imry Ma length $\xi (h_0) $
for $ d \leq 2 $ is much larger than the size of domains $ L(t) $ in the
time region of interest. Therefore, we shall deal with coarsening,
irrespective of dimensionality. 
Using the split~(\ref{4.10}) and following the argument of the previous
Section we have again $\langle s_i(t) \rangle _h 
= \overline{\sigma _i(t)}^h$. Expanding up to linear order we generalize
(\ref{3.9}) by writing
\begin{equation}
\overline{\sigma _i(t)} ^h =  \sum _j
\chi_B (\vert i-j \vert,t- t _w) h _j 
+ \chi_I (\vert i-j \vert,t,t _w) h _j
+ {\cal O}(h ^2)
\label{4.111}
\end{equation}
where the integrated response function has been separated into the sum
of two contributions. The first one accounts for the change in the 
magnetization in the bulk of domains and, due to the separation of
the time scales of fast and slow relaxation, is TTI. In other words, this
contribution ignores the existence of interfaces and therefore under
all respects is the same as the integrated response analyzed in the
fast relaxation process of the previous Section, i.e.
$\chi_B(t-t_w) = \chi_{\alpha}(t-t_w)$. Instead, the second
contribution accounts for the extra response due to the
existence of interfaces and is not TTI. Inserting in~(\ref{3.8}), in 
place of~(\ref{3.10}) we now have
\be
M(t,t_w) = T\chi_B(t-t_w) + T\chi_I(t,t_w).
\label{4.112}
\ee
By definition, the bulk contribution obeys FDT and therefore, 
following the discussion of the previous Section, is related to
the autocorrelation function by~(\ref{3.21}), with the
difference that now the region $C < m_T^2$ is not unphysical.

For what concerns the interface contribution, in the first time
regime with $ t-t_w \ll t_w $ interfaces can be regarded as static and
 it is quite reasonable to take $ \chi _I $ proportional
 to the interface density 
$\chi _I (t,t_w) \simeq \rho _I (t_w) \sim L^{-1}(t_w)$.
The question is what happens in the aging regime $t-t_w \gg t_w$
dominated by interface
motion. The usual argument~\cite{Berthier99,FMPP,Parisi99} 
is that $\chi_I$ keeps on being proportional 
to the interface density
\begin{equation}
\chi _I(t,t_w) \sim \rho _I(t)
\label{4.20}
\end{equation}
and therefore eventually disappears like $ L^{-1}(t) $.
If so, it is clear that by taking $ t_w $ large
enough the interface contribution can be made negligible with respect
to the bulk contribution, leaving~(\ref{3.21}) to account for the
relation between the whole response and the autocorrelation function.
However, as explained in the Introduction and as we shall see later on, 
the interface response function
turns out to have more structure than what~(\ref{4.20}) allows for.
In particular, there is a dependence on space dimensionality which cannot
be accounted for by~(\ref{4.20}).

\section{Statics from Dynamics} \label{FMPP}

Let us now come to the problem of detecting the structure of the equilibrium
state from the properties of the linear response function in the 
off-equilibrium regime. This requires, first of all, the generalization
of FDT out of equilibrium. As we have seen in Section 3  when FDT holds
the time dependence of $\chi(t-t_w)$ is absorbed into the dependence
on $C(t-t_w)$. In the FDT generalization proposed by Cugliandolo 
and Kurchan~\cite{Cugliandolo93}, as stated in the Introduction, 
this holds also in the aging regime postulating that for large
$t_w$
\be
\chi(t,t_w) =\chi[C(t,t_w)].
\label{5.1}
\ee
In order to establish the connection between the FDR~(\ref{1.4}) 
and the static
properties, FMPP have considered the general case
in which at the time $t_w$ the hamiltonian is changed into
${\cal H}_{J}[s]={\cal H}_0[s] - {\cal H}_p[s]$
with a perturbation of the form
${\cal H}_p[s] = \sum_{i_{1}<..<i_{p}} J_{i_{1}..i_{p}}s_{i_1}..s_{i_p}$
where the couplings $J_{i_{1}..i_{p}}$ are independent gaussian variables.
By considering the behavior of the expectation 
$E_{J}[\langle {\cal H}_{p}(t) \rangle _J]$
they have derived the connection between the FDR and 
the overlap probability
function of the equilibrium state. Here, for simplicity, we reproduce 
the main steps
of the argument in the particular case of $p=1$,
when the perturbation ${\cal H}_p$ reduces to the form~(\ref{2.13}),
referring to~\cite{FMPP} for the treatment with arbitrary $p$.

The expectation entering in the definition of the staggered magnetization
can be written as
\be
E_{h}[\langle {\cal H}_{1}(t) \rangle_h] = 
E_h \sum_{\omega} \rho_{h}(\omega,t){\cal H}_1(\omega)
\label{5.5}
\ee
where $ \rho_{h}(\omega,t)$ is the probability distribution evolving with 
the hamiltonian~(\ref{2.12}).
Using the fact that $h_{i}$
are independent gaussian variables and integrating by parts
\be
E_h\left [\langle {\cal H}_1(t)\rangle _h\right ]= h_0^2 E_h \left[ \sum_{i}
\frac{\partial}{\partial h_{i}} 
\left( \sum_{\omega}  \rho_{h}(\omega,t) s_{i} \right) \right].
\label{5.6}
\ee
The same quantity can be evaluated dynamically in the Martin-Siggia-Rose
formalism~\cite{FMPP} and, assuming that~(\ref{5.1}) holds, one finds
\bea
E_h\left [\langle {\cal H}_1(t)\rangle _h\right ]
&=& \frac{h_0^2}{T}N E_h \Bigg [ 1 - C_h(t,t_w) X_h(C_h(t,t_w))\nonumber \\
&-& \left .\int_{C_h(t,t_w)}^{1} dq \frac{dX_h(q)}
{dq} q \right]
\label{5.7}
\eea
where $X_h$ and $C_h$ are, respectively, the FDR and the autocorrelation
function in the perturbed system. Taking the 
$t \rightarrow \infty$ limit and using $\lim_{t \rightarrow \infty}
C_h(t,t_w)=0$, from~(\ref{5.6}) and~(\ref{5.7}) follows
\be
E_h \left[ \sum_{i} \frac{\partial}{\partial h_{i}}
\left< s_{i}\right>_{h,\infty} \right] =
 \frac{N}{T}  \left [ 1-E_h \left ( \int_{0}^{1} dq \frac{dX_h(q)}
{dq} q \right ) \right]
\label{5.8}
\ee
where $\langle \cdot \rangle_{h,\infty}$ denotes the average with the 
probability distribution
\be
\lim_{t \rightarrow \infty} \rho_{h}(\omega,t) = \rho_{h}(\omega,\infty).
\label{5.9}
\ee

Therefore, what we have up to this point is that under  
assumption~(\ref{5.1}) there exists a relation between the susceptibility in
the state~(\ref{5.9}) and the first moment of the FDR in the perturbed system.
In the general case one has the 
same relation between the generalized
susceptibility with respect to $J_{i_1 ... i_p}$ 
and the $p$-th moment of the FDR. In order to go further
on, more must be known about the properties of $\rho_{h}(\omega,\infty)$. 
In the context considered by FMPP $\rho _h (\omega,\infty)$ coincides with
the perturbed Gibbs state~(\ref{2.14}),
from which follows
\be
 \frac{\partial}{\partial h_{i}}
\left< s_{i}\right>_{G,h} = \frac{1}{T} \left[ 1 -
\left< s_{i}\right>_{G,h}^2 \right].
\label{5.10}
\ee
Inserting this in the left hand side of~(\ref{5.8}) and using~(\ref{2.61})
one finds
\be
E_h \int dq P_{G,h}(q)q=E_h\int _0 ^1 dq \frac {dX_h(q)}{dq}q.
\label{11}
\ee
From this and from similar relations for the higher moments one may
establish the identity $P_{G,h}(q)=\frac {d}{dq}X_h(q)$
which yields $\tilde P_G (q)=\frac {d}{dq}\tilde X(q)$
where $\tilde P_G(q)=\lim _{h\to 0 }P_{G,h} (q)$
and $\tilde X(q)=\lim _{h\to 0 }X_h (q)$.
Eventually, after establishing under what conditions
$\tilde P_G (q)$ and $\tilde X(q)$ may be identified, respectively, with the
overlap function $P_G (q)$ of the unperturbed Gibbs state and 
with the FDR $X(q)$
of the unperturbed dynamics, one has the connection between the unperturbed 
statics and dynamics in the form
\be
P_G(q)=\frac{dX(q)}{dq}.
\label{5.2.2}
\ee
Here, we call the attention on the fact that to establish~(\ref{11})
it is essential that~(\ref{5.10}) holds in order to use~(\ref{2.61}).
Furthermore, the derivative with respect to $h_i$ in the left hand side 
of~(\ref{5.8}) acts according to how the RF enters in the asymptotic
state~(\ref{5.9}). Therefore, if instead of 
reaching the Gibbs state~(\ref{2.14})
the asymptotic state $\rho _h (\omega, \infty)$ coincides with the 
state~(\ref{3.7}), in place of~(\ref{11}) one finds
\be
E_h\int dq P_{\alpha ,h}(q)q=E_h\int _0 ^1 dq \frac{dX_h(q)}{dq}q
\label{5.4.1}
\ee
where $P_{\alpha ,h}(q)$ is the overlap function of the perturbed
pure state. Hence, following through the argument illustrated above, in 
place of~(\ref{5.2.2}) one concludes that the unperturbed FDR is related
to the overlap function of the pure unperturbed state
\be
P_\alpha (q)=\frac{dX(q)}{dq}.
\label{5.4.2}
\ee

In summary, the static information contained in the FDR depends on how
the perturbation ${\cal H}_p$ enters in the asymptotic state~(\ref{5.9}).
We must now see in what form the scheme applies to the phase ordering process.
Suppose that the interface response function $\chi _I(t,t_w)$ can be neglected
in~(\ref{4.112}) for $t_w$ sufficiently large. 
Then, as explained in the previous Section, 
assumption~(\ref{5.1}) is verified and $M(C)$ obeys~(\ref{3.21})
leading to~(\ref{3.225}) which coincides with~(\ref{5.4.2}).
This is confirmed by numerical simulations on the Ising model for
$d\geq 2$~\cite{Barrat98,Parisi99} which show evidence 
for convergence toward the 
form~(\ref{3.21}) in the parametric plot of $M$ versus $C$ as
$t_w$ becomes large.  Therefore, a behavior of the
type~(\ref{3.21}) is the signature that 
the interface contribution
to the response function is negligible and the phase ordering process
behaves as the fast relaxation process of Section~\ref{fastrelax}.

\section{Ising model $d=1$} \label{ising1}

In this Section and the next one we analyze the linear response in the
off-equilibrium dynamics of the Ising model beginning from the one
dimensional case where analytical results are available.

The system is defined by the hamiltonian with nearest
neighbor interaction
${\cal H}(\omega) =-J\sum _ {i}s_is_{i+1}$,
where $J>0$ is the ferromagnetic coupling.
From equilibrium statistical mechanics we know that 
the equilibrium correlation function behaves as
$\langle s_is_j\rangle _G=e^{-\frac{\vert i-j\vert}{\xi _T}}$
and the correlation length is given by
$\xi _T = -\left [\ln \tanh \left (\frac{J}{T}\right ) \right ]^{-1}$.
Therefore ergodicity is broken for $K=J/T=\infty $, which requires either
$T=0$ for $J<\infty $ or $J=\infty $ and $T$ arbitrary.
Solving dynamics with $K=\infty $,
the two time correlation function
obeys the form~(\ref{4.3}) where only the aging 
contribution~(\ref{4.4})
is present with $m^2_T=1$.
The explicit form of the autocorrelation function is given by~(\ref{1.22}).
The reason for the absence of the TTI contribution $C_{ps} (\vert i-j \vert,
t-t')$  is clear since with $K=\infty $ 
the ordering component $\sigma _i=\pm 1$ coincides
with $s_i$ and $\psi _i$ in~(\ref{4.10}) vanishes identically. 
Namely, in the quench
with $K=\infty $ domains are formed, phase space motion takes place
over $\Omega _b$, as demonstrated by the aging behavior of the correlation
function, but thermal fluctuations are absent within domains leading
to the absence of the TTI contribution in the autocorrelation function.

It is interesting to consider also what happens in the quench with $K<\infty $.
In this case there is no ergodicity breaking. 
Solving dynamics one finds the generalized scaling form~\cite{Bray89}
$C(\vert i-j \vert,t,t')=F\left ( \frac {\vert i-j \vert }{L(t')},
\frac{L(t)}{L(t')},\frac{L(t')}{\xi _T}\right )$
with the limiting behaviors
\bea
F& &\left ( \frac {\vert i-j \vert }{L(t')},
\frac{L(t)}{L(t')},\frac{L(t')}{\xi _T}\right )= \nonumber \\
 & & \left \{ \begin{array}{ll}
                   F_{st}\left ( \frac {\vert i-j \vert }{\xi _T},
                   \frac{t-t'}{\tau}\right )
                              & \mbox{, \quad for $\frac{t'}{\tau} \gg 1$} \\
                   F_{ag}\left ( \frac {\vert i-j \vert }{L(t')},
                   \frac{L(t)}{L(t')}\right )
                            & \mbox{, \quad for $\frac {t-t'}{\tau} \ll 1$
and $\frac{t'}{\tau} \ll 1$}
                   \end{array}
              \right .
\label{6.8}
\eea
where the equilibration time $\tau$ is defined by~(\ref{2.500}) with $z=2$.
The meaning of~(\ref{6.8})
is, first of all, that after a finite time $\tau $
equilibrium is reached. Hence, taking $t'/\tau >1$ one 
observes the TTI behavior pertaining to the
stationary dynamics in the equilibrium state. However, if $\tau$ although
finite is sufficiently large to allow for $t/\tau \ll 1$, then in the
time regime $(t-t')\ll \tau$ one observes the same behavior as in the
$K=\infty $ case. This can be understood considering that for $K$ large
$\tau \sim e^{4K}$ is the characteristic time needed to overturn
one spin originally aligned with both of its neighbors. Then, taking
$t' \ll \tau$ means that one starts to look into the system when domains
are still much smaller than $\xi _T$. Immediately after and as long
as $t< \tau$ growth continues as in the $K=\infty $ case, i.e. without
thermal fluctuations within domains. Thermal fluctuations do come into play
only for $t\geq \tau$. When this happens, the creation of defects by 
thermal fluctuations balances the losses due to interface annihilation
and leads to a halt in domain growth and to the establishment of thermal
equilibrium.

Let us see what happens upon applying the RF
at the time $t_w>0$. Recall that the master equation for the system
evolving with the Glauber spin flip dynamics is given by
\be
\frac{\partial \rho \left (\omega ,t\right )}{\partial t}=
\sum _i \left [w(-s_i)\rho \left (R_i\omega,t\right )-w(s_i)
\rho \left (\omega,t\right )
\right ]
\label{6.10}
\ee
where $R_i\omega$ is the configuration $\omega$ with the $i$-th spin
reversed and $w(s_i)$ is the transition rate from $\omega$ to $R_i\omega$
given by
\be
w(s_i)=\frac{1}{2}(1-H^{int}_is_i)(1-H^{ext}_is_i)
\label{6.11}
\ee
where $H^{int}_i=\frac{\gamma }{2}(s_{i-1}+s_{i+1})$, 
$H^{ext}_i=\tanh \left (\frac {h_i}{T}\right )$
and $\gamma =\tanh (2K)$.
Taking $K=\infty $, let us first consider the behavior of~(\ref{6.11})
before switching on the external field, in the interval $0<t<t_w$.
Since $\gamma =1$, we have $H^{int}_i=1$ if $s_{i-1}s_{i+1}=1$ and
$H^{int}_i =0$ if $s_{i-1}s_{i+1}=-1$. In the latter case the spin $s_i$
is at the interface between two domains of opposite sign with
probability $1/2$ per unit time to flip or not to flip. 
Conversely, in the former
case the spin flips with probability 1 if it is not aligned with its
neighbors, while it does not flip with probability 1 in the opposite case,
when it belongs to the bulk of a domain. As a consequence the only dynamics
in the system is the unbiased random walk of interfaces, leading to the growth
law $L(t)\sim t^{1/2}$.

\begin{figure}
\resizebox{0.5\textwidth}{!}{%
  \includegraphics{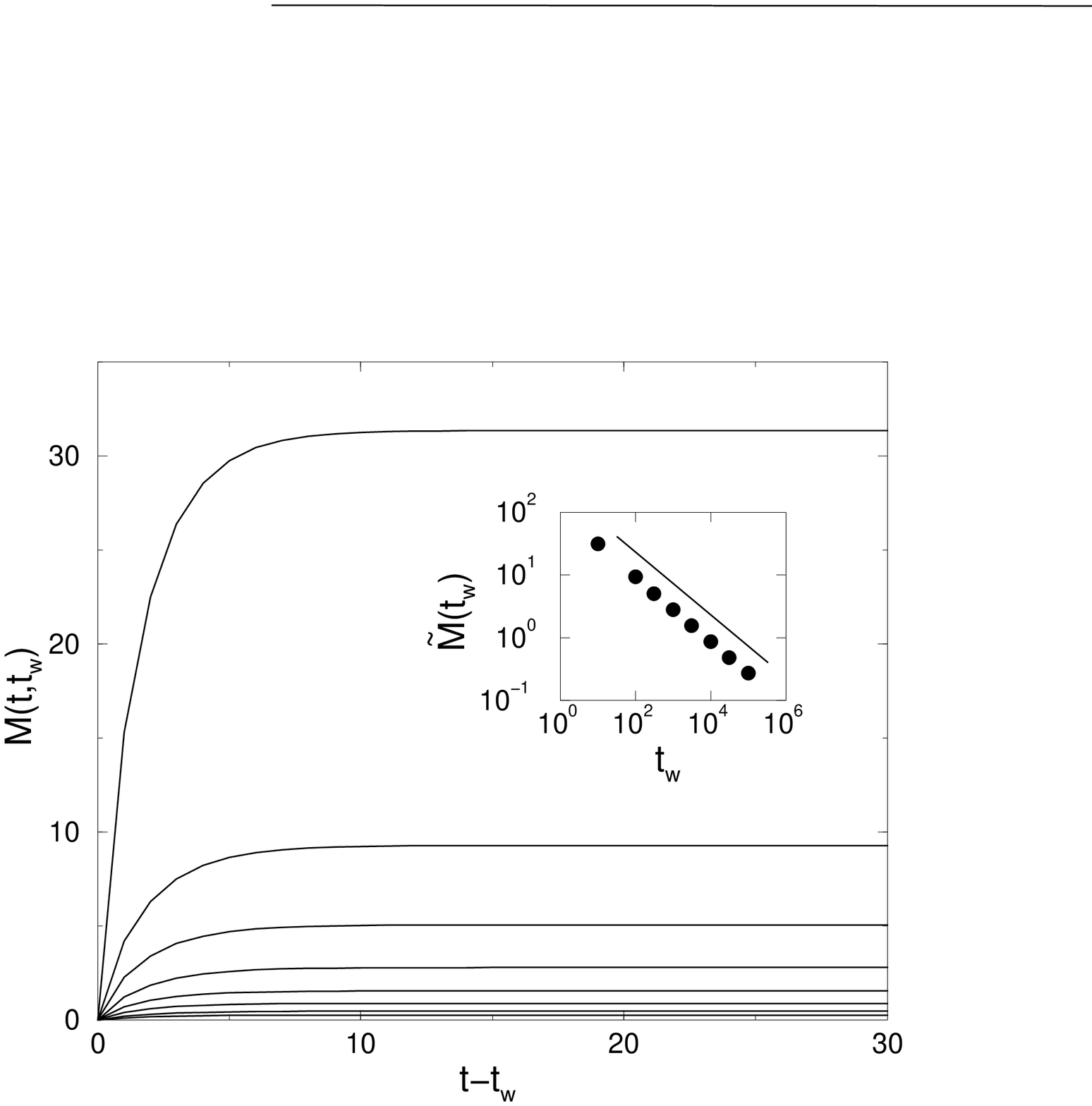}
}
\caption{$M(t,t_w)$ for the $d=1$
Ising model with $J=1$ quenched to $T=0$ for different waiting times
($10,10^2,3.16\cdot 10^2,10^3,3.162\cdot 10^3,10^4, 
3.162 \cdot 10^4,10^5$ from top to bottom).
In the inset, the plateau value $\widetilde M (t_w)$ is plotted
against $t_w$. The solid line is the $t_w^{-1/2}$ behavior.}
\label{largeh}
\end{figure}

After switching on the RF we are interested in the behavior
of the staggered magnetization~(\ref{3.8}), which is
now convenient to regard as the correlation function between the external
field and magnetization. Right at the start $M(t_w,t_w)=0$, since at
$t=t_w$ the RF and configurations are uncorrelated.
However, for $t>t_w$ the transition rate~(\ref{6.11}) is modified
by the factor involving $H^{ext}_i$ which introduces a bias in the flips
at the interfaces in favor of the local field, while bulk flips
remain forbidden. Accordingly, $M(t,t_w)$ grows positive revealing
that spin configurations tend to correlate with the field. However,
a substantial difference arises in the two ways to produce the
limit $K=\infty $. If $J<\infty $ and $T=0$, $M(t,t_w)$
rises from zero to a plateau value $\widetilde{M}$ (Fig.~\ref{largeh}) 
within a microscopic 
time $t_0$ and $\widetilde{M}$ depends on $t_w$ according to 
(inset of Fig.~\ref{largeh})
\be
\widetilde{M}(t_w)\sim L^{-1}(t_w).
\label{6.15}
\ee

The reason for this is that after switching on the field the motion of
interfaces continues for a short time until pinning takes place
when the two opposite spins making up the defect at the interface
fall into a defect of the same sign in the field configuration.
At that point, since $H^{ext}_i=\mbox {sign} \, h_i$,
the second factor in the
right hand side of~(\ref{6.11}) vanishes and the interface does not move
anymore. This explains reaching the plateau and the dependence~(\ref{6.15})
of $\widetilde{M}$ on $t_w$, since the contribution to the
staggered magnetization comes only from the spins at the interfaces and
goes like the interface density. Furthermore, 
since $T=0$ implies $h_0/T=\infty$
there is no way to access the linear response regime, no matter how small
the external field is chosen.
Conversely, if the $K=\infty $ limit is obtained taking $J=\infty $ and
with no restrictions on the temperature, while the unperturbed dynamics
remains the same, interesting behavior is obtained with RF since 
{\it i}) $T>0$ allows to overcome
pinning of the interfaces letting also
the bulk of domains to participate in the correlation of spin configurations
with the RF and {\it ii}) the condition $h_0/T\ll 1$ can be realized
making accessible the linear response regime.

In the following we will concentrate in the linear response regime with
$J=\infty $ and $h_0/T \ll 1$. In this case the staggered magnetization
is given by~(\ref{1.20}).
The first observation, comparing with the general form~(\ref{4.112}),
is that the TTI bulk contribution is missing, as expected 
from the above discussion on the absence of
thermal fluctuations when $K=\infty $. Hence, the
result~(\ref{1.20}) is entirely due to the second contribution
in~(\ref{4.112}), which however is totally 
different from the behavior~(\ref{4.20})
which one would expect on the basis of a straightforward interface 
contribution.
Rather than decreasing at large time, here $M(t,t_w)$ displays a correlation
of the spin configurations with the RF which grows with time,
until reaching the finite limit~(\ref{1.21}) as $t \rightarrow \infty$ 
(Fig.~\ref{isinguno}).
Having excluded a correlation effect due to thermal fluctuations or to 
spin polarization at the interfaces, the increase in the
correlation can only be due to the fact that interface motion
is driven by the field. As we shall now see, the field driven
mechanism, without modifying the growth law $L(t)\sim t^{1/2}$,
induces large scale domain drift in order to optimize the position
of the bulk of domains with respect to the RF
configuration. It must be stressed that although involving
the bulk of domains, this contribution to the staggered magnetization
has nothing to do with the bulk response function coming from thermal 
fluctuations, which are now absent.

An insight on how the field driven mechanism works comes from the behavior
of~(\ref{1.20}) for $t-t_w \ll t_w$ from which we find
\be
\chi _I(t,t_w)=\frac{1}{T \pi }\left [ \frac{2(t-t_w)}{t_w}\right ] 
^{\frac{1}{2}}.
\label{6.18}
\ee
Since in this time regime the system can be regarded as a set of non interacting
interfaces and the density of interfaces $\rho _I(t)$ at the time $t\simeq t_w $
is given by $L^{-1}(t_w)$, we can rewrite~(\ref{6.18}) in the form
\be
\chi _I (t,t_w)=\rho _I (t)\chi _{eff}(t,t_w)
\label{6.19}
\ee
where
\be
\chi _{eff}(t,t_w)\sim (t-t_w)^{\frac{1}{2}}
\label{6.20}
\ee
is the effective response associated to a {\it single} interface.
Looking next to the large time behavior for $t-t_w \gg t_w$, from~(\ref{1.20})
follows $T\chi _I(t,t_w)=1/\sqrt 2 -{\cal O}(t/t_w)^{-1/2}$ which implies
that the effective single interface response follows the behavior~(\ref{6.20})
also for large time.

In order to check on this interpretation we have computed analytically
the behavior of $\chi _I(t,t_w)$ when in the system there is a single
interface. This is done by preparing the system at $t=0$ in a
configuration $\omega$ containing only one interface at the origin,
for instance taking $s_i=1$ for $i\leq 0$ and $s_i=-1$ for $i>0$.
The computation of the response function can be carried out 
exactly (see Appendix A) yielding
\be
\chi _{sing}(t-t_w)\sim (t-t_w)^{\frac{1}{2}}
\label{6.22}
\ee
which substantiates the above analysis. This unexpected result makes it
clear that the interface response is not simply due to the polarization
of the paramagnetic interfacial spins, but is a much more complex effect.
At a generic time $t$ the interface has explored a region of order
$(t-t_w)^{1/2}$ and energy can only be released by reducing the contribution
to ${\cal H}_1(\omega) $ coming from the visited region. This can be achieved 
if the interface motion produces
a large scale optimization of the position of domains
with respect to the random field. 

In summary, from the analysis of the linear response function in the quench
of the $d=1$ Ising model with $K=\infty $, we have uncovered a
new and non trivial behavior different from the pattern outlined
in Section~\ref{slow}, which is the one usually expected for coarsening
systems. The role of the bulk and interface terms 
in~(\ref{4.112}) is reversed, the response being dominated by the latter
one with all the new features illustrated above.

\section{Ising model $D>1$} \label{$d>1$}

As stated in Section 4, for the Ising model in two and
three dimensions there is numerical evidence that 
as $ t_w $ becomes large the staggered magnetization 
displays the structure ~(\ref{3.21}). For this to happen,
the interface contribution in~(\ref{4.112}) must vanish and only the
bulk contribution must be left over. However, the exact analysis of the
previous Section in the one dimensional case is a serious warning that
the interface contribution might not disappear so easily as the 
argument~(\ref{4.20}) could make believe. 
Therefore, a careful analysis of the
interface contribution is needed to find out whether the field driven
mechanism of interface motion is at work also in higher space
dimensions.
In order to do so it is necessary to give an anambiguous definition
of which degrees of
feedom must be cosidered interfacial.
In particular, it is necessary to make clear whether the flip of a 
spin in the interior of a domain belongs to a bulk fluctuation or 
makes a new interface.
The definition we adopt is the following. Consider two configurations
$\omega_{I+B}$ and $\omega_{I}$ 
evolving from the same initial condition with the same 
thermal history, under the influence of the same
external field, if present. While $\omega_{I+B}$ evolves with the usual
Glauber dynamics, $\omega_{I}$ is subjected to the 
restriction that flips of bulk spins are forbidden. 
A bulk spin is defined as being aligned with all
its nearest neighbours. Then, all the spins surrounding topological defects
in $\omega_{I}$ are considered interfacial spins. On the other hand
$\omega_{I+B}$ contains defects which can be either associated to
interfaces or to bulk fluctuations, the latter  being determined by comparison
with $\omega_{I}$. 
On the basis of this definition we have measured the interface
response function $\chi _I(t,t_w)$ by simulating the evolution of
the ferromagnetic Ising model with nearest neighbor interaction
for $d=1,2,3,4$ without bulk flips and starting
from the high temperature uncorrelated initial condition~(\ref{4.2}).
We have included the simulation of the $d=1$ case, for which the exact
analytical results of the previous Section are available, in order to
have a check on the numerical procedure.

\begin{figure}
\resizebox{0.5\textwidth}{!}{%
  \includegraphics{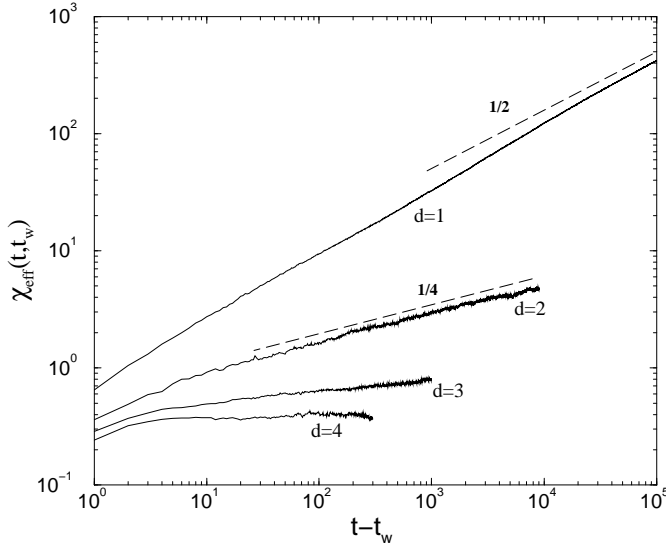}
}
\caption{$\chi _{eff}(t,t_w)$ in the Ising
model without spin flips in the bulk.
For $d=1,2,3,4$, the temperature, waiting time and linear system size 
of the simulations are 
$T=0.48, 2.2,3.3,4.4$, $t_w=10^3,10^3,10^2,10$ and
$L=10^6,512,200,42$ with $J=1$ and averages 
over $170,6045,114,922$ realizations.
The dashed lines are power laws 
with the corresponding exponent $\alpha $.
For $d=3$ the curve is well fitted by 
$T\chi _{eff}(t,t_w)=0.33+0.066\cdot log(t-t_w)$.}

\label{chi_eff}
\end{figure}

For the effective response
associated to a single interface $\chi _{eff}(t,t_w)$ defined by~(\ref{6.19}),
we have obtained (Fig.~\ref{chi_eff}) the asymptotic behavior
\be
\chi _{eff}(t,t_w) \sim (t-t_w)^\alpha
\label{7.1}
\ee
where the numerical values of  $\alpha$ are consistent with
\begin{equation}
\alpha = \left \{ \begin{array}{ll}
                          \frac{3-d}{4}   & \mbox{for $d<3$} \\
                          0               & \mbox{for $d>3$}.
                       \end{array}
             \right .
\label{7.2}
\end{equation}
For $d=3$ a power law fit yields $\alpha=0.03$. A fit of the same quality
is obtained with the logarithmic form
$T \chi _{eff}(t,t_w)=0.33+0.066\cdot log(t-t_w)$. 

\begin{figure}
\resizebox{0.5\textwidth}{!}{%
  \includegraphics{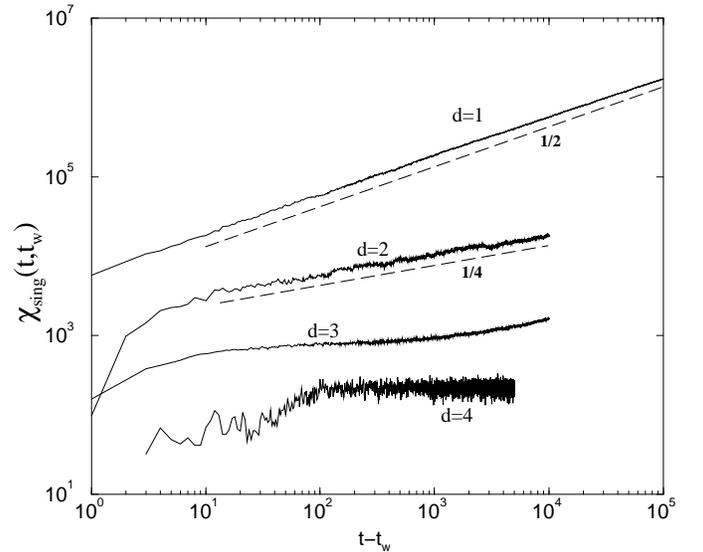}
}
\caption{The single interface response, 
obtained from simulations
without flips in the bulk and with an initial condition containing
one single flat interface. 
For $d=1,2,3,4$ temperatures  
of the simulations are 
$T=0.48, 2.2,3.3,4.4$, 
with $J=1$, $t_w=0$ and averages 
over $151600,19810,69891,3537$ realizations.
The dashed lines are power laws 
with the corresponding exponent $\alpha $.}
\label{chi_sing}
\end{figure}

In order to check to what extent $\chi _{eff}(t,t_w)$ is related to a
single interface response, we have performed another set of simulations
without flips in the bulk, starting with an initial condition containing
one straight spanning interface in the middle of the system. 
The results of simulations are shown in Fig.~\ref{chi_sing} and indeed
the data reproduce quite well the behavior of $\chi _{eff}(t,t_w)$, except
for $d=3$. In this case the logarithmic behavior found for
$\chi _{eff}(t,t_w)$ is followed up to a certain time, beyond which the
response speeds up considerably. The analysis of this behavior, for
which we do not have an adequate explanation, requires numerical
investigation at much larger times and goes beyond the scope of the
present work.

In summary, comparing Fig.~\ref{chi_eff} with Fig.~\ref{chi_sing}, 
the identification
of $\chi _{eff}(t,t_w)$ with the response associated to a single
interface $\chi _{sing}(t,t_w)$ is on the whole well founded. 
We must now extract the
meaning of the overall behavior as dimensionality is varied. The main feature is that the power law growth~(\ref{7.1}) weakens as $d$
rises from $d=1$ to $d=3$ and disappears above $d=3$. The explanation
of this behavior can be conjectured recalling that in the unperturbed system
interfaces perform an unbiased random walk in $d=1$, while are curvature 
driven for $d \geq 2$.
In the perturbed system in $d=1$, as we have seen in the previous Section,
interfaces are field driven. This mechanism operates also in higher 
dimensions, except that it enters in competition with the curvature mechanism. 
The effect of
the curvature becomes comparatively more important as $d$ increases due
to the increasing coordination number. According to our simulations
$d_c =3$ is the critical value of the dimensionality, 
such that for $d> d_c$ the 
field driven mechanism is ineffective and interface motion is dominated by
the curvature yielding $\alpha=0$. Therefore, for $d> d_c$, interfaces respond
only through the polarization of the interfacial spins.
This response saturates to the asymptotic value over a microscopic time
scale (curves for $d=4$ in Fig.\ref{chi_eff} and Fig.\ref{chi_sing}). 
Conversely, for $d< d_c$, the field driven mechanism competes
with the curvature yielding $\alpha>0$ and this competition gets more effective
as dimensionality is lowered. Finally, as the limit $d=1$ is reached 
from above, the curvature 
mechanism disappears and the effectiveness of the field driven mechanism 
becomes complete yielding $\alpha=1/2$.

\section{Continuous spin model} \label{continuous}

The discussion of the previous Section makes clear that dimensionality
plays a crucial role in determining the relative importance of the bulk
and interface response. In order to clarify further this point, in this
Section we present an analytical calculation of $\chi _I (t,t_w)$ 
in the framework of
continuous spins which allows to vary dimensionality freely. 
The approximations involved in what follows are too crude for an 
accurate quantitative reproduction of the results of the simulations.
Nonetheless, even in this form, the continuous model is quite useful
to capture the overall qualitative picture.

The treatment of phase ordering with a continuous, scalar and non conserved
order parameter $\phi (\vec x,t)$ is usually based~\cite{Bray94}
on the time dependent Ginzburg-Landau equation
\be
\frac {\partial \phi }{\partial t} = \nabla ^2 \phi +r \phi
-g\phi ^3 +\eta (\vec x,t)
\label{8.1}
\ee
where $r$ and $g$ are positive constants and $\eta (\vec x,t)$
is a gaussian white noise with expectations
\be
\left \{ \begin{array}{ll}
        \langle \eta (\vec x,t)\rangle =0 \\
  \langle \eta (\vec x',t')\eta (\vec x,t)\rangle=2T\delta (t-t')
\delta (\vec x  - \vec x').
        \end{array}
        \right .
\label{8.2}
\end{equation}
The infinite temperature initial
condition is imposed taking
$\phi _0(\vec x)=\phi(\vec x,t=0)$ as an additional source of
noise gaussianly distributed with expectations
\be
\left \{ \begin{array}{ll}
        \overline {\phi _0(\vec x)} =0 \\
        \overline { \phi _0(\vec x) \phi _0(\vec x')}=\Delta \delta (\vec x-\vec x').
        \end{array}
        \right .
\label{8.3}
\end{equation}
In past work on phase ordering kinetics most of the effort has been devoted to the
study of the scaling properties of the equal time correlation function,
that is, in the present language, to the study of $C_{ag}(\mid i-j \mid,t=t')$
in~(\ref{4.3}) and~(\ref{4.4}). For this purpose, thermal fluctuations
are usually neglected eliminating the thermal noise in~(\ref{8.1}). Then, 
one deals with the equation
\be
\frac{\partial \phi}{\partial t}=\nabla ^2 \phi +r\phi-g\phi ^3
\label{8.4}
\ee
where the only source of noise is the initial condition $\phi _0(\vec x)$.

From the analytical point of view, one of the most successful tools of
investigation of this problem has been the gaussian auxiliary field (GAF)
approximation which goes back to the pioneering work of Ohta, Jasnow
and Kawasaki~\cite{Ohta82,Bray94}. 
The method is suited to study the late stage, after local
equilibrium within domains has been achieved. In the case of Eq.~(\ref{8.4})
this means that locally the order parameter must sit at the bottom of
either one of the two degenerate minima of the potential satisfying
\be
\phi (\vec x,t)=\pm m_0
\label{8.5}
\ee
where $m_0=\sqrt {r/g}$ is the $T=0$ equilibrium value of the order parameter.
The idea at the basis of the GAF approximation is that~(\ref{8.5}) can be
implemented through a non linear transformation on an auxiliary field
$u(\vec x,t)$ over which perturbative methods can be applied. Different
versions of the approximation correspond to different realizations of the
non linear transformation. Here we take the transformation of the
Kawasaki, Yalabik and Gunton~\cite{Kawasaki78} type
\be
\phi (\vec x,t)=\frac{u(\vec x,t)}{\sqrt{1+\frac{u^2(\vec x,t)}{m_0^2}}}.
\label{8.6}
\ee
Then, if the non linearity of $u(\vec x,t)$
is mild, unbounded growth is allowed eventually yielding
$\phi (\vec x,t)=m_0 sign [u(\vec x,t)]$
which enforces the requirement~(\ref{8.5}). In order to actually carry
out computations, one has to solve the dynamics of $u(\vec x,t)$
induced by~(\ref{8.4}) via~(\ref{8.6}), as we shall do below.

After this brief survey of the GAF method, let us go back to the
equation of motion~(\ref{8.1}) including thermal fluctuations. A systematic 
treatment of this problem based on the Martin-Siggia-Rose formalism and
on the split of the order parameter into ordering and fluctuating
components was worked out in Ref.~\cite{MVZ}. Here, we follow the same
idea working directly with the equation of motion. Let us split the
order parameter as in~(\ref{4.10})
\be
\phi (\vec x,t)=\psi (\vec x,t)+\sigma (\vec x,t)
\label{8.8}
\ee
with the aim of separating the fast thermal fluctuations from the slow
ordering component. Inserting~(\ref{8.8}) in~(\ref{8.1}) we find
\bea
\frac {\partial \psi}{\partial t}+\frac{\partial \sigma}{\partial t}&=&
\nabla ^2 \psi +\nabla ^2 \sigma +r\psi +r\sigma-g\psi ^3- 
\nonumber \\         
                                                                    & & 
3g\psi ^2\sigma
-3g\psi \sigma ^2-3g\sigma ^3 +\eta
\label{8.9}
\eea
and let us decouple $\psi $ from $\sigma $ replacing the mixed terms
by $3g\psi ^2\sigma \rightarrow 3g \langle \psi ^2 \rangle \sigma$ and
$3g\psi \sigma ^2 \rightarrow 3g\psi \overline{ \sigma ^2}$.
Furthermore, let us assume that $T$ is sufficiently lower than
$T_c$ to justify the self-consistent linearization
$\psi ^3 \rightarrow 3 \langle \psi ^2 \rangle \psi$.
Stipulating that $\psi$ is driven by the thermal noise and that $\sigma$
is driven by the noise in the initial condition, we obtain the pair
of equations
\be
\frac {\partial \psi }{\partial t}=\nabla ^2 \psi +
\left [ r-3g \langle \psi ^2(\vec x,t) \rangle
-3g \overline {\sigma ^2 (\vec x,t)} \right ] \psi+\eta
\label{8.10}
\ee
and 
\be
\frac {\partial \sigma }{\partial t}=\nabla ^2 \sigma +\left [
r-3g\langle \psi ^2(\vec x,t)\rangle \right ] \sigma-g\sigma ^3
\label{8.11}
\ee
with the initial conditions $\psi (\vec x,t=0)=0$ and
 $\sigma (\vec x,t=0)=\phi _0(\vec x)$. Let us make the assumption, 
to be verified {\it a posteriori}, that $\psi $ is the
fast variable with relaxation time $\tau $. Defining
$r_{eq}=r-3g \langle \psi ^2\rangle _{eq}$ and making the additional 
assumption that within the same time scale $\sigma (\vec x,t)$ reaches
local equilibrium with
$\overline {\sigma ^2 (\vec x,t)}=m^2_T=\frac{r_{eq}}{g}$,
for $t> \tau$ in place of~(\ref{8.10}) and~(\ref{8.11}) we may write
\begin{eqnarray}
\frac {\partial \psi }{\partial t} &=&
            \nabla ^2 \psi -\xi _T^{-2}\psi+\eta (\vec x,t)
            \label{8.12}\\
\frac {\partial \sigma }{\partial t} &=&
            \nabla ^2 \sigma +r_{eq}\sigma -g\sigma ^3
            \label{8.13}
\end{eqnarray}
where the equilibrium correlation length $\xi _T$ is given by
\be
\xi_T ^{-2}=2r_{eq}.
\label{8.14}
\ee

From~(\ref{8.8}) follows that the autocorrelation function is given by
the sum of two contributions as in~(\ref{4.3})
$C(t,t')=C_{ps}(t-t')+C_{ag}(t,t')$
with the TTI piece
\be
C_{ps}(t-t')=\langle \psi (\vec x,t)\psi (\vec x,t')\rangle=
\langle \psi ^2 (\vec x)\rangle _{eq} e^{-\frac{t-t'}{\xi_T^2}}
\label{8.17}
\ee
and the aging contribution
\be
C_{ag}(t,t')=\overline {\sigma (\vec x,t)\sigma (\vec x,t')}.
\label{8.18}
\ee
The latter one can be computed from~(\ref{8.13}) using the GAF
approximation with the non linear transformation of the
type~(\ref{8.6}) in which $m_0^2$ is replaced by $m^2_T$
\be
\sigma (\vec x,t)=\frac {u(\vec x,t)}{\sqrt {1+\frac{u^2(\vec x,t)}{m_T^2}}}.
\label{8.19}
\ee
From~(\ref{8.17}) indeed follows that $\psi (\vec x,t)$ describes the fast
equilibrating thermal fluctuations with the characteristic time
$\tau=\xi^2_T$.

Consider, next, the effect on the ordering component  
of an RF with expectations analogue to~(\ref{2.17})
\be
\left \{ \begin{array}{ll}
        E_h[h(\vec x)]=0 \\
 	E_h[h(\vec x)h(\vec x')]=h_0^2\delta (\vec x-\vec x').
        \end{array}
        \right .
\label{8.20}
\end{equation}
For $t > t_w$ the
equation of motion~(\ref{8.13}) is modified into
\be
\frac{\partial \sigma }{\partial t}=\nabla ^2 \sigma +r_{eq}\sigma 
-g\sigma ^3+h(\vec x)
\label{8.21}
\ee
while~(\ref{8.12}) for $\psi (\vec{x},t)$ remains unaltered.
In order to generalize the GAF approximation, notice that the external
field affects the transformation~(\ref{8.19}) in two ways:
{\it i}) through the auxiliary field $u(\vec x,t)$ and {\it ii}) by shifting
the saturation value $\pm m_T$ of domains. Therefore, we separate
a bulk and an interface term 
writing $\sigma(\vec x,t) = \sigma _B(\vec x,t)+\sigma _I (\vec x,t)$
where
\be
\sigma _B(\vec x,t)=\int d\vec x'\chi _B(\vec x-\vec x',t-t_w)h(\vec x')
\label{8.22}
\ee
and
\be
\sigma _I(\vec x,t)=\frac {u_h(\vec x,t)}
{\sqrt {1+\frac{u_h^2(\vec x,t)}{m^2_T}}}.
\label{8.23}
\ee
The latter one is constructed to account only for the effect
of the external field on the interface motion, by keeping the saturation
value at the unperturbed level $\pm m_T$, while the former takes care
of the remaining perturbation on the bulk of domains. Hence, for the
staggered magnetization the decomposition~(\ref{4.112}) applies
where, according to the discussion of Section \ref{slow}, $\chi _B$ obeys FDT
and is therefore related to the autocorrelation function by~(\ref{3.21}).
Here we are interested in the interface contribution
\be
\chi _I (t,t_w)=\frac{1}{h_0^2}E_h\left [\overline {\sigma _I (\vec x,t)}
h(\vec x)\right ]
\label{8.24}
\ee
and in order to compute this quantity let us go back to~(\ref{8.21}).
Since we want to extract the dependence of $u_h(\vec x,t)$ on the RF
up to first order, after substituting $\sigma =\sigma _B+\sigma _I$
and keeping into account that $\sigma _B$ is a first order quantity we
obtain
\be
\frac {\partial \sigma _I}{\partial t}=\nabla ^2\sigma _I+r_{eq}\sigma _I
-g\sigma ^3_I+ h(\vec x)
\label{8.27}
\ee
where the effect of $\sigma_B$ goes into a redefinition of the variance 
$h_0^2$ of the RF, which will be neglected in the following.
Substituting~(\ref{8.23}) for $\sigma _I(\vec x,t)$ and dropping
the subscripts $h$, the equation of motion for the auxiliary field
is given by
\be
\frac{\partial u}{\partial t}=\nabla ^2 u+r_{eq}u+
\frac{\sigma ''(u)}{\sigma '(u)}\left ( \nabla u\right )^2+
\frac{h(\vec x)}{\sigma '(u)}
\label{8.29}
\ee
where
\be
\sigma '(u)=\left [ 1+\frac{u^2}{m^2_T}\right ]^{-\frac{3}{2}}
\label{8.29.1}
\ee
and
\be
\sigma ''(u)=-3\frac {u}{m_T^2}
\left [ 1+\frac{u^2}{m^2_T}\right ]^{-\frac{5}{2}}.
\label{8.29.2}
\ee
Performing, next, a mean field approximation by keeping only the
lowest order non linear contribution in each term and linearizing
self-consistently, after Fourier transforming over space we find
\be
\frac{\partial u(\vec k,t)}{\partial t}=-[k^2+D(t)]u(\vec k,t)+
h(\vec k)
\label{8.29.3}
\ee
where
\be
D(t)=-r_{eq}+\frac{3}{m^2_T}\overline {(\nabla u)^2}.
\label{8.29.4}
\ee

Defining the linear response function by
\be
R(\vec k,t,t')=\frac{Y(t',0)}{Y(t,0)}e^{-k^2(t-t')}
\label{8.29.5}
\ee
with
\be
Y(t,0)=e^{\int _0^t ds D(s)}
\label{8.29.6}
\ee
the formal solution of~(\ref{8.29.3}) reads
\be
u(\vec k,t)=R(\vec k,t,0)u(\vec k,0)+\chi _u(\vec k,t,t_w) h(\vec k)
\label{8.29.7}
\ee
where $\chi _u(\vec k,t,t_w)=\int _{t_w} ^t dt' R(\vec k,t,t')$
is the integrated response function of the auxiliary field.
Carrying out the self-consistent computation of $D(t)$ (Appendix B),
the large time behavior of $Y(t,0)$ is given by
$Y(t,0)=$ 

\noindent $a\left ( t+\frac{1}{2\Lambda ^2}\right ) ^{-\frac{d+2}{4}}$
where $\Lambda $ is a momentum cutoff and
$a=\left [ \frac {3\Delta d}
{4r_{eq}m^2_T(8\pi)^{\frac{d}{2}}}
\right ]^{\frac{1}{2}}$.
Inserting in~(\ref{8.29.5}), from
$\chi _u(t,t_w)=$ \\
$(2\pi)^{-d}\int d\vec k \chi _u(\vec k,t,t_w)$
and $t_w \gg 1/\Lambda ^2$ follows
\be
\chi _u(t,t_w)=(4\pi)^{-\frac{d}{2}}t^{\frac{d+2}{4}}
\int _{t_w} ^t dt' t'^{-\frac{d+2}{4}}\left (t-t'+\frac{1}{\Lambda ^2}\right )
^{-\frac{d}{2}}.
\label{8.32}
\ee
Similarly, for the unperturbed autocorrelation function of the $u$ field we
find (Appendix B)
\be
\overline{u(\vec x,t) u(\vec x,t')} =
\frac{ \Delta \left [ \left (t+ \frac {1}{2\Lambda^2} \right )
\left (t'+ \frac {1}{2\Lambda^2} \right ) \right ]^{\frac{d+2}{4}}}
{a^2 (4\pi)^{d/2}  \left (t+t'+ \frac {1}{\Lambda^2} \right )^{d/2}}.
\label{8.32.1}
\ee
Now, in order to compute~(\ref{8.24}) we make a further mean field
approximation by replacing~(\ref{8.23}) with
\be
\sigma _I(\vec x,t)=m_T \frac {u_h(\vec x,t)}{\sqrt {S(t)}}
\label{8.25}
\ee
where $S(t)=\overline {u^2(\vec x,t)}$
is an unperturbed average. This gives
\be
\chi _I(t,t_w)=
\frac{m_T}{\sqrt{S(t)}} \chi _u(t,t_w)
\label{8.26}
\ee
and computing $S(t)$ from~(\ref{8.32.1}) we get
\be
S(t)=b\left (t+\frac{1}{2\Lambda ^2} \right )
\label{8.33}
\ee
with $b=\frac {4r_{eq}m_T^2}{3d}$ and
\be
\chi _I(t,t_w)=At^{\frac{1-d}{2}}F\left ( \frac{t_w}{t},\frac{t_0}{t}\right )
\label{8.34}
\ee
where
\be
F\left ( \frac{t_w}{t},\frac{t_0}{t}\right )=
\int _{\frac{t_w}{t}}^1 dx x^{-\frac{d+2}{4}}
\left ( 1+\frac{t_0}{t}-x\right )^{-\frac{d}{2}}.
\label{8.35}
\ee
Here, $t_0=\Lambda  ^{-2}$ is a microscopic time and
$A=\left [ \frac{3d}{4r_{eq} (4\pi)^d }\right ]^{\frac{1}{2}}$.

\begin{figure}
\resizebox{0.5\textwidth}{!}{%
  \includegraphics{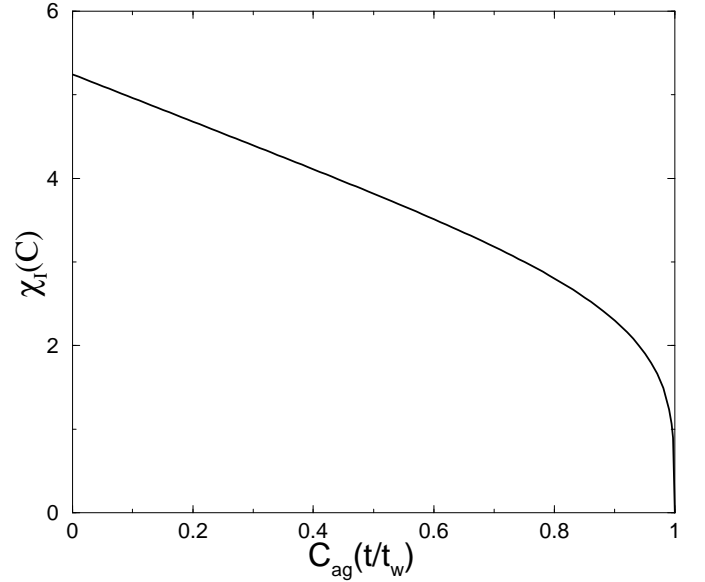}
}
\caption{Parametric plot of $\chi _I$ against $C_{ag}$ in the
continuous spin model.}
\label{gaf_d1}
\end{figure}

Next, we may use the form~(\ref{8.25}) of the transformation to compute also
the aging contribution~(\ref{8.18})  to the autocorrelation 
function obtaining (Appendix B)
\be
C_{ag}(t/t_w) = m_T^2 \left (\frac{t_w}{t} \right )^{d/4}
\left ( \frac{1}{2}+\frac{t_w}{2t} \right )^{-d/2}.
\label{8.36.1}
\ee
For $d=1$ the time ratio $t_w/t$ can be eliminated between~(\ref{8.34})
and~(\ref{8.36.1}) yielding a parametric plot (Fig.~\ref{gaf_d1}) of the
response function versus $C$ qualitatively similar to the one of 
the $d=1$ Ising model in Fig.~\ref{isinguno}. In particular,
in the large time limit we find the counterpart of~(\ref{1.21})
\be
\lim_{t \rightarrow \infty} \chi_I (t,t_w) = AF(0,0) =
\sqrt{\frac{3}{2r_{eq}}}\left [ \frac{\Gamma(1/4)}{2\pi} \right ]^2.
\label{8.36.2}
\ee

\begin{figure}
\resizebox{0.5\textwidth}{!}{%
  \includegraphics{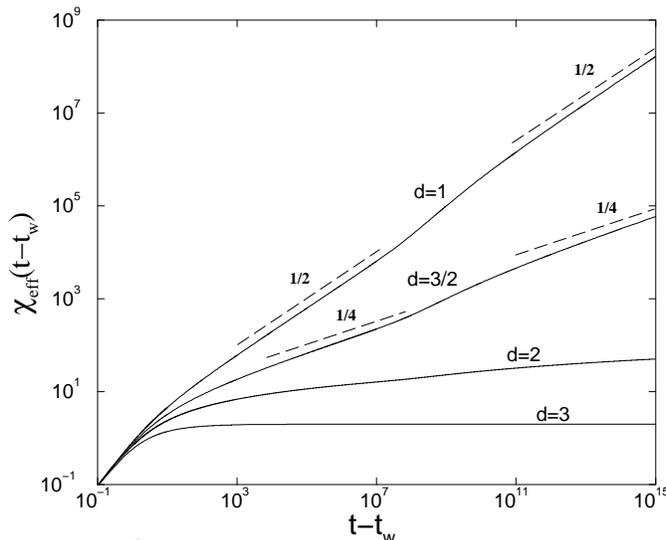}
}
\caption{$\chi _{eff}(t,t_w)$ in the continuous spin
model with $t_w=10^8$. 
The dashed lines are power laws 
with the corresponding exponent $\alpha $.}
\label{gaf_alld}
\end{figure}

The interesting point now is to extract the behavior of the effective
response function $\chi _{eff}$ associated to the single interface
and defined  by~(\ref{6.19}). In the short time regime $t-t_w\ll t_w$
from~(\ref{8.34}) follows
\be
\chi _{eff}(t-t_w)=\frac{2\Lambda ^{d-2}}{2-d}\left [ \left (
\frac{t-t_w}{t_0}+1\right )^{\frac{2-d}{2}}-1\right ]
\label{8.37a}
\ee
which for $t_0\ll t-t_w \ll t_w$ yields
\be
\chi _{eff}(t-t_w)=\left \{ \begin{array}{ll}
		   \frac {2\Lambda ^{d-2}}{d-2} & \mbox{, \quad for $d>2$} \\
                   \log \left (\frac{t-t_w}{t_0}\right ) & \mbox{, \quad for $d=2$} \\
                   \frac{2\Lambda ^{d-2}}{2-d}\left (\frac {t-t_w}{t_0}
                   	\right )^{\frac{2-d}{2}} & \mbox{' \quad for $d<2$}.
        \end{array}
        \right .
\label{8.37}
\end{equation}
A similar behavior is obtained also in the large time regime
$t-t_w\gg t_w$
\be
\chi _{eff}(t-t_w)=\left \{ \begin{array}{ll}
		   \frac {2\Lambda ^{d-2}}{d-2}A & \mbox{, \quad for $d>2$} \\
                   4A\log \left (\frac{t}{t_w}\right ) & \mbox{, \quad for $d=2$} \\
                   AF(0,0)t^{\frac{2-d}{2}} & \mbox{, \quad for $d<2$}.
        \end{array}
        \right .
\label{8.38}
\end{equation}
Therefore, apart from a change in the prefactor taking place about
$t-t_w\sim t_w$, from~(\ref{8.37}) and~(\ref{8.38}) follows that both
for short and large time $\chi _{eff}$ obeys a power law as in~(\ref{7.1})
where, however, now
\be
\alpha=\left \{ \begin{array}{ll}
		   \frac {2-d}{2} & \mbox{, \quad for $d<2$} \\
                   0 & \mbox{, \quad for $d>2$}
        \end{array}
        \right .
\label{8.39bis}
\end{equation}
and there is logarithmic growth for $d=2$.
The full time dependence of $\chi _{eff}(t,t_w)$ obtained 
from the numerical computation of~(\ref{8.34}) 
for different values of $d$ is displayed in Fig.~\ref{gaf_alld}.
Comparing Fig.~\ref{chi_eff} and Fig.~\ref{gaf_alld}, the common
features may be
summarized stating that in both cases $\chi _{eff}$ obeys the
power law~(\ref{7.1}) and that there exists a critical value of the 
dimensionality
$d_c$ such that the exponent $\alpha$ is zero for $d \ge d_c$ 
with logarithmic growth at $d=d_c$. For $d <d_c$ the exponent $\alpha$
grows positive with decreasing dimensionality reaching
the final value $\alpha =1/2$ at $d=1$.
The meaning of the critical dimensionality in relation to the growth
mechanism has been discussed in the previous Section.
The difference with the case of Ising spins is that now we have 
$d_c=2$ in place of $d_c=3$. This tells that, although qualitatively
correct, the mean field approximation developed above is not accurate
enough to account quantitatively for the competition between the field driven
and curvature driven growth mechanisms. For instance, we find a domain growth
law $L(t)\sim t^{1/2}$ also in $d=1$, while the one dimensional
continuous model is known to have logarithmic growth law~\cite{Nagai86}.
Despite these shortcomings, the model reproduces the gross features
of the response function as dimensionality is varied.

\section{Conclusions}

\begin{figure}
\resizebox{0.5\textwidth}{!}{%
  \includegraphics{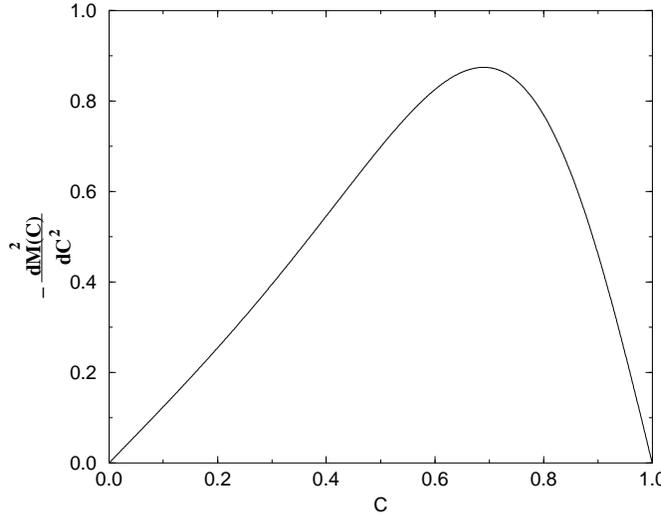}
}
\caption{$-\frac{ d^2 {\cal M} (C)}{d C^2}$
for the $d=1$ Ising model with $J=\infty $.}
\label{p_di_q}
\end{figure}

In this paper we have studied the behavior of the response function
in non disordered coarsening systems under variation of space 
dimensionality. The results obtained are instructive on the
applicability of the FMPP theorem in general. In order to clarify
this point, let us go back to the form~(\ref{1.1}) of the 
staggered magnetization
\be
M(C,t_w) = M_{st}(C) + t_w^{-a}{\cal M}(C)
\label{9.1}
\ee
where we have used~(\ref{1.10}).
As we have emphasized, the above pattern in the response function
reveals the existence of slow and fast degrees of freedom with
widely separated time scales. When $P(q)$ is extracted from~(\ref{9.1})
two basically different cases must be distinguished. If $a\neq 0$,
as it is the case for coarsening systems with $d>1$, 
the slow degrees of freedom
for large $t_w$ make a negligible contribution and the relevant
information comes only from $M_{st}(C)$ yielding $P_{st}(q)=
\delta (q-q_{EA})$, where $q_{EA}$ is the Edwards-Anderson order
parameter ($q_{EA}=m_T^2$). Conversely, if $a=0$ one obtains
an additional contribution due to the slow degrees of freedom
\be
P(q)=P_{st}(q)+P_{ag}(q)
\label{9.2}
\ee
where $P_{ag}(q)=-d^2 {\cal M}(C)/dC^2 \vert _{C=q}$.
This non trivial contribution appears in glassy systems and
reproduces the expected pattern of replica symmetry breaking
of the equilibrium state~\cite{Parisi99}. However, this
quantity appears also in the $d=1$ Ising model with (Fig.~\ref{p_di_q})
\be
\left. -\frac{ d^2 {\cal M} (C)}{d C^2} \right ]_{C=q} =
\frac{\pi cos(\frac{\pi q}{2}) sin(\frac{\pi q}{2})}
{[2- sin(\frac{\pi q}{2})]^2}
\label{9.3}
\ee
and in this case it is not related to the equilibrium state.
The general question then is: if during some relaxation process
the response function takes the form~(\ref{9.1}) with $a=0$, under what 
conditions ${\cal M}(C)$ contains information on the equilibrium state.
The preceding analysis for coarsening systems suggests that
the answer has to do with one of
the hypothesis in the theorem, which requires the system eventually
to equilibrate, and with the mechanism of slow relaxation.
In glassy systems the time evolution proceeds toward equilibrium through 
decays of metastable states~\cite{FMPP}. 
This may take very long, but eventually
all degrees of freedom, including the slow ones, will equilibrate.
The case of coarsening, instead, is qualitatively different.
Slow relaxation is not due to decay of metastable states, there are no
activated processes. Rather, there is a smooth reduction of defect
energy, as motion in phase space takes place over the border,
where the slow degrees of freedom do reduce in number but 
never equilibrate. Hence, in this case,
${\cal M}(C)$ is a property of an intrinsically out of equilibrium dynamics
with no connection to any property of the equilibrium state.

As a simple illustration, let us briefly consider the case of free diffusion
\be
\frac{\partial \phi}{\partial t}=\nabla ^2 \phi +\eta
\label{9.4}
\ee
which demonstrates quite well the existence of non equilibrating
degrees of freedom whose visibility depends on space dimensionality.
In Fourier space~(\ref{9.4}) takes the form
\be
\frac{\partial \phi(\vec k,t)}{\partial t}=-k^2 \phi (\vec k,t) 
+\eta (\vec k,t)
\label{9.5}
\ee
which shows that all the modes with $\vec k \neq 0$ equilibrate
while the $\vec k=0$ mode executes Brownian motion and therefore
never equilibrates~\cite{Cugliandolo94}. The linear response
function is given by $R(\vec k,t,t')=\exp [-k^2(t-t')]$.
Integrating over $\vec k$ and over time, for the integrated 
response function
\be
\chi (t,t_w)=\int _{t_w}^t dt' \int \frac{d^d k}{(2\pi)^d}
e^{-k^2(t-t')}
\ee
we find for large time the same pattern of behavior as in~(\ref{8.37})
\be
\chi (t,t_w)=\left \{ \begin{array}{ll}
		   (4\pi )^{-\frac{d}{2}}\frac {2\Lambda ^{d-2}}{d-2} & \mbox{, \quad for $d>2$} \\
                   (4\pi )^{-\frac{d}{2}} \log \left (\frac{t-t_w}{t_0}\right ) & \mbox{, \quad for $d=2$} \\
                   (4\pi )^{-\frac{d}{2}} \frac{2\Lambda ^{d-2}}{2-d}\left (\frac {t-t_w}{t_0}
                   	\right )^{\frac{2-d}{2}} & \mbox{' \quad for $d<2$}.
        \end{array}
        \right .
\label{9.6}
\end{equation}
A similar behavior is displayed by the equal time correlation function
$C(t,t)=\langle \phi ^2(\vec x,t)\rangle $ for large time
\be
C(t,t)=\left \{ \begin{array}{ll}
		   T(4\pi )^{-\frac{d}{2}}\frac {2\Lambda ^{d-2}}{d-2} & \mbox{, \quad for $d>2$} \\
                   T(4\pi )^{-\frac{d}{2}} \log \left (\frac{2t}{t_0}\right ) & \mbox{, \quad for $d=2$} \\
                   T(4\pi )^{-\frac{d}{2}} \frac{2\Lambda ^{d-2}}{2-d}\left (\frac {2t}{t_0}
                   	\right )^{\frac{2-d}{2}} & \mbox{' \quad for $d<2$}.
        \end{array}
        \right .
\label{9.7}
\end{equation}
Finally, from~(\ref{9.6}) and~(\ref{9.7}) follows
\be
\lim _{t\to \infty}\frac{T\chi (t,t_w)}{C(t,t)}=\left \{ \begin{array}{ll}
		   1&                 \mbox{, \quad for $d>2$} \\
                   1&                 \mbox{, \quad for $d=2$} \\
                   2^{\frac{d-2}{2}}& \mbox{' \quad for $d<2$}.
        \end{array}
        \right .
\label{9.8}
\end{equation}
These results expose the basic mechanism responsible of the behavior
of the response function. When looking in $\vec x$ space all the
$\vec k$ modes are mixed together and the existence of one of them 
which does not equilibrate is hidden by the density of states as long
as $d>2$, but cannot be canceled for $d\leq 2$ and becomes more evident
the lower is the dimensionality. In particular, (\ref{9.8}) shows that
the out of equilibrium $\vec k=0$ mode does not prevent the equilibrium
FDT to be asymptotically satisfied for $d>2$ and also for $d=2$,
but for $d<2$ a deviation from equilibrium FDT appears which is
increasingly important as $d\to 1$. Interestingly enough, for
$d=1$ one recovers $\lim _{t\to \infty} T\chi (t,t_w)=C(t,t)/\sqrt 2$
as in~(\ref{1.21}) for the $d=1$ Ising model, recalling that for
Ising spins $C(t,t)=1$.

\vskip 1cm

{\small This work was partially
supported by MURST PRIN-2000 and by the European TMR Network-Fractals Contract
No. FMRXCT980183. F.C. acknowledges support by
INFM PRA-HOP 1999.}

\vskip 1cm

\appendix {Appendix A} \label{appendix}

In order to compute $\chi _{sing}(t,t_w)$ in the $d=1$ Ising model let us
first recall that in the exact solution of the model~\cite{Glauber63}
the two times and the equal times correlation
functions are related by
\begin{equation}
C(\mid i-j \mid,t,t^{\prime}) = \sum_l C(\mid j-l \mid,t^{\prime},t^{\prime})
F_{i-l}(t-t^{\prime}).
\label{dodici}
\end{equation}
where $F_{i-m}(t-t^{\prime})= e^{-(t-t^{\prime})}
I_{i-m}[\gamma (t-t^{\prime})]$ and $I_{n}(x)$ are the Bessel
functions of imaginary argument. From this follows
\bea
{\partial \over \partial t^{\prime}} &C&(\mid i-j\mid ,t,t^{\prime})+
{\partial \over \partial t} C(\mid i-j\mid ,t,t^{\prime}) =\nonumber \\
&\sum_l& {dC(\mid j-l \mid ,t^{\prime } ,t^{\prime } ) \over dt^{\prime}}
  F_{i-l}(t-t^{\prime}).
\label{ventisei}
\eea 
The linear response function 
\be
R_{i,j}(t,t^{\prime}) = \left ( {\delta <s_{i}(t)>_{h} \over
\delta h_{j}(t^{\prime})} \right)_{h=0}
\label{diciannove}
\end{equation}
can be cast in the form~\cite{Lippiello00}
\begin{equation}
R_{i,j}(t,t^{\prime } )
=\frac{2}{T}F_{i-j}(t-t^{\prime } )\langle w(s^{\prime }_j) \rangle
\label{1eq}
\end{equation}
where $ w(s^{\prime }_j)$ is the unperturbed transition rate.
Rewriting the right hand side as 
\bea
\frac{2}{T}\langle w(s^{\prime }_j)&\rangle& F _{i-j}=
\frac{1}{T}\left \{ \sum _m \langle s^{\prime } _js^{\prime }_m
 [w(s^{\prime } _j)+w(s^{\prime } _m)]\rangle F_{i-m}\right . \nonumber \\ 
&-&\left . \sum _{m \neq j} \langle
s^{\prime } _js^{\prime }_m [w(s^{\prime }_j)+w(s^{\prime }_m)]\rangle 
F _{i-m} \right \}
\label{1eq1}
\eea
and using Glauber evolution equation for 
$C(\mid i-j \mid ,t^{\prime } ,t^{\prime }) $
\bea
& & \frac{dC(\mid i-m \mid,t^{\prime } ,t^{\prime }) }
{dt^{\prime } }= \nonumber \\
& & \left \{ \begin{array}{ll} 
                 -2\langle s^{\prime } _i s^{\prime} _m 
\left [ w(s^{\prime } _i)+w(s^{\prime } _m)\right ]\rangle    
                                & \mbox{for $m \neq i$} \\
                  0             & \mbox{for $m=i$}
                \end{array}
      \right .
\label{aeq}
\eea
from~(\ref{1eq}) and~(\ref{1eq1}) one obtains
\bea
TR _{i,j}(t,t')&=& \frac{1}{2} \sum _m \frac{dC(\mid j-m \mid,t^{\prime },
t^{\prime })}{dt^{\prime } } 
F _{i-m}(t-t^{\prime } ) \nonumber \\
               &+& B_ {i,j}(t,t^{\prime})
\label{cicciolino}
\eea
with 
\begin{equation}
B _{i,j}(t,t^{\prime})= \sum _m \langle s^{\prime } _js^{\prime }_m\left 
[w(s ^{\prime } _j)+w(s^{\prime } _m)\right ] 
\rangle F_{i-m}(t-t^{\prime } ). 
\label{bij}
\end{equation}
Next, inserting~(\ref{ventisei}) in~(\ref{cicciolino}) we obtain 
\bea
TR _{i,j}(t,t')&=& \frac{1}{2} 
\left [ {\partial \over \partial t^{\prime}} C(\mid i-j \mid ,t,t^{\prime}) 
\right .\nonumber \\
&+& \left .{\partial \over \partial t} C(\mid i-j \mid ,t,t^{\prime})\right ]
+B_ {i,j}(t,t^{\prime }) 
\label {nuova1}
\eea
taking $i=j$ and summing over $i$ this gives
\bea
T\sum _i R _{i,i}(t,t')&=& \frac{1}{2} \sum _i
\left [ {\partial \over \partial t^{\prime}} C(i,t,t^{\prime}) 
\right . \nonumber \\
&+& \left . {\partial \over \partial t} C(i,t,t^{\prime})\right ]
+B(t,t^{\prime }) 
\label {nuova4}
\eea
where $B(t,t^{\prime })=\sum _i B_ {i,i}(t,t^{\prime })$ and
$C(i,t,t^{\prime })=C(\mid i-j \mid =0,t,t^{\prime })$
is the autocorrelation function. In the general case of
absence of space translation invariance this quantity
depends on the site $i$. 
Furthermore, from~(\ref{bij})
\begin{eqnarray}
B(t,t^{\prime } ) & = & \sum _{i,m} \langle s^{\prime } _is^{\prime }
 _m\left [ w(s^{\prime }_i)+w(s^{\prime }_m)\right ]
\rangle F_{i-m}(t-t^{\prime } ) \nonumber \\ 
& = & \sum _i \langle s^{\prime }_i w(s^{\prime }_i)\sum _m s^{\prime }
 _m\rangle F_{i-m}(t-t^{\prime }) \nonumber \\
&+& \sum _m \langle s^{\prime }_m w(s^{\prime } _m)
\sum _i s^{\prime } _i\rangle 
F_{i-m}(t-t^{\prime })
\label{pippo1}
\end{eqnarray}
using the parity of Bessel function $ F _x(z) = F _{-x}(z) $ and 
the result of Ref.~\cite{Glauber63}
\be
\sum _{\omega ^\prime} \rho (\omega' t^\prime \mid \omega t)s_i
=\sum _l s_l ^\prime F_{i-l}(t-t^\prime )
\ee
we find
\begin{eqnarray}
&B&(t,t^{\prime })=2\sum _i \langle s^{\prime }_i w(s^{\prime }_i)
\sum _m s^{\prime } _m\rangle F_{i-m}(t-t^{\prime } )\nonumber\\
& = & 2\sum _i \sum _{\omega, \omega ^{\prime } }s^{\prime }_is_iw
(s^{\prime }_i)\rho (\omega ^{\prime } ,t^{\prime } )\rho (\omega ,t\mid 
\omega ^{\prime } ,t^{\prime }).
\label{4beq}
\end{eqnarray}
Since the conditional probability obeys the master equation~\cite{Glauber63}
\begin{equation} 
{\partial \over \partial t} \rho (\omega ,t\mid \omega ^{\prime }  ,t^{\prime } )
  = -\sum _m s _m \sum _{s^{\prime \prime } _m} s^{\prime \prime }  _m
w(s^{\prime \prime }  _m) \rho (\omega ,t\mid \omega ^{\prime }  ,t^{\prime } )
\label{booo}
\end{equation}
this gives
\begin{equation}
B(t,t^{\prime })= 
-\sum _i {\partial \over \partial t}C(i;t,t^{\prime } )
\label{4eq}
\end{equation}
and putting this result in~(\ref{nuova4}) we finally obtain
\begin{equation}
T \sum _i R _{i,i}(t,t^\prime ) = \frac{1}{2} \sum _i
\left [ {\partial \over \partial t^{\prime}} C(i;t,t^{\prime}) 
-{\partial \over \partial t} C(i;t,t^{\prime})\right ].
\label{nuova5}
\end{equation}
Up to this point the results we have obtained are fully general. Let
us now specialize to the case of the initial condition with a single
interface, e.g. $\omega(t=0) = [ s_i =1 $ for  
$i \leq 0, s_i =-1 $ for $i>0]$.
Furthermore, if we take $J=\infty$ also the evolving configuration
will contain a single interface, namely
$\omega(t) = [ s_i =1$ for $i \leq l(t), 
s_i =-1 $ for $ i>l(t)]$
where $l(t)$ is the position of the interface at the time $t$. If we
consider two configurations at the times $t,t'$ we have
\be
\sum_i s_i(t)s_i(t')= N - 2\mid l(t)-l(t')\mid
\label{II172}
\ee
where $N$ is the total number of spins. Taking the thermal average
\be
\sum_i C(i,t,t') = N -2x(t-t')
\label{II173}
\ee
where $x(t-t')= \langle \mid l(t)-l(t')\mid \rangle$ is the average
of the absolute value of the displacement of the interface. Since
this quantity is TTI we may write ${\cal C}(t-t')= \sum_i C(i,t,t')$
and inserting in~(\ref{nuova5}) we find
\be
T{\cal R}(t-t') = \frac{d{\cal C}(t-t')}{dt'}
\label{II174}
\ee
where ${\cal R}(t-t')=\sum_i R_{i,i}(t,t')$. 

\noindent Defining $\chi _{sing}=(1/N)\int _{t_w}^t {\cal R}(t-t')dt'$
we get
\be
NT\chi _{sing}(t-t_w) = [{\cal C}(t=t_w)- {\cal C}(t-t_w)] = 2x(t-t_w)
\label{II175}
\ee
which yields~(\ref{6.22}) keeping into account that
$x(t-t_w) \sim (t-t_w)^{\frac{1}{2}}$.

\appendix{Appendix B} \label{appendix2}

Taking for the initial expectations of the auxiliary field
\be
\left \{ \begin{array}{ll}
                   \overline {u(\vec k,0)}=0 \\
                   \overline {u(\vec k_1,0)u(\vec k_2,0)}=
                   	(2\pi )^d\Delta \delta (\vec k_1 +\vec k_2)
                   \end{array}
              \right .
\label{a2.1}
\ee
and using~(\ref{8.29.7}) the unperturbed averages are given by
\bea
I(t)&=&\overline {(\nabla u)^2}=\Delta \int \frac {d^dk}{(2\pi )^d}
k^2 R^2(\vec k,t,0)e^{-\frac{k^2}{\Lambda ^2}}\nonumber \\
&=& \frac {\Delta}{Y^2(t,0)} \int \frac {d^dk}{(2\pi )^d}
k^2 e^{-2k^2(t+\frac{1}{2\Lambda ^2})}
\label{a2.2}
\eea
and
\bea
& &\overline {u(\vec x,t)u(\vec x,t')}= 
\Delta \int \frac {d^dk}{(2\pi )^d}
R(\vec k,t,0) R(\vec k,t',0) \nonumber \\
&=& \frac {\Delta}{Y(t,0)Y(t',0)}
\int \frac {d^dk}{(2\pi )^d}
e^{-k^2(t+ t'+\frac{1}{\Lambda ^2})}
\label{a2.2.1}
\eea
where $\Lambda $ is the momentum cutoff.
Next, using
\be
\int \frac {d^dk}{(2\pi )^d}
e^{-2k^2x}=(8\pi x)^{-\frac{d}{2}}
\label{a2.4}
\ee
and
\be
\int \frac {d^dk}{(2\pi )^d}
k^2e^{-2k^2x}= 2\pi d (8\pi x)^{-\frac{d+2}{2}}
\label{a2.5}
\ee
we have
\be
I(t)=\frac {2\pi d \Delta }{Y^2(t,0)} (8\pi )^{-\frac{d+2}{2}}
\left (t+\frac {1}{2\Lambda ^2} \right )^{-\frac{d+2}{2}}.
\label{a2.8}
\ee
From~(\ref{8.29.6})
\be
\frac {dY^2 (t,0)}{dt}=2D(t)Y^2(t,0)=
-2r_{eq}Y^2(t,0)+\frac{6}{m^2_T}I(t)Y^2(t,0)
\label{a2.9}
\ee
and, neglecting the time derivative on the left hand side, for large
time we find
\be
Y^2(t,0)=a^2\left ( t+\frac{1}{2\Lambda ^2} \right )^{-\frac{d+2}{2}}
\label{a2.10}
\ee
with $a^2=\frac{3\Delta d }
{4r_{eq}m^2_T(8\pi)^{\frac{d}{2}}}$.
Inserting in~(\ref{a2.2.1}) we have
\be
\overline{u(\vec x,t) u(\vec x,t')} =
\frac{ \Delta \left [ \left (t+ \frac {1}{2\Lambda^2} \right )
\left (t'+ \frac {1}{2\Lambda^2} \right ) \right ]^{\frac{d+2}{4}}}
{a^2 (4\pi)^{d/2}  \left (t+t'+ \frac {1}{\Lambda^2} \right )^{d/2}}.
\label{a2.11.1}
\ee
Using~(\ref{8.25}) the aging contribution to the correlation function
is given by
\be
C_{ag}(t,t') = \overline{\sigma(\vec x,t)\sigma(\vec x,t')}=
m_T^2 \frac{\overline{u(\vec x,t) u(\vec x,t')}}
{\sqrt{S(t)S(t')}}
\label{a2.11.2}
\ee
and inserting~(\ref{8.33}) and~(\ref{a2.11.1}) this gives
\be
C_{ag}(t,t') = m_T^2  
\frac{\left [ \left (1+ \frac{t_0}{2t}\right )
\left (\frac{t'}{t}+ \frac{t_0}{2t}\right ) \right ]^{d/4}}
{\left (\frac{1}{2}+\frac{t_0}{2t}+ \frac{t'}{2t}\right )^{d/2}}
\label{a2.11.3}
\ee
which reduces to~(\ref{8.36.1}) for $t_0 /t \rightarrow 0$.


\begin{thebibliography}{000}


\bibitem{Bouchaud97}
J.P.Bouchaud, L.F.Cugliandolo, J.Kurchan and M.Mezard, in
\textit{Spin Glasses and Random Fields} edited by A.P.Young
(World Scientific, Singapore, 1997).

\bibitem{MVZ}
G.F.Mazenko, O.T.Valls and M.Zannetti, Phys.Rev. B \textbf{38}, (1988) 520.

\bibitem{Berthier99}
L.Berthier, J.L.Barrat and J.Kurchan, Eur.Phys.J. B \textbf{11}, (1999) 635.

\bibitem{Franz00}
S.Franz and M.A.Virasoro, J.Phys. A \textbf{33}, (2000) 891.

\bibitem{Cugliandolo93}
L.F.Cugliandolo and J.Kurchan, Phys.Rev.Lett. \textbf{71}, (1993) 173;
Philos.Mag. {\bf 71}, 501 (1995); J.Phys. A \textbf{27}, (1994) 5749.

\bibitem{FMPP}
S.Franz, M.Mezard, G.Parisi and L.Peliti, Phys.Rev.Lett. {\bf 81},
1758 (1998); J.Stat.Phys. {\bf 97}, 459 (1998).

\bibitem{Mezard87}
M.Mezard, G.Parisi and M.A.Virasoro, \textit{Spin Glass Theory and Beyond}
(World Scientific, Singapore, 1987).

\bibitem{Barrat98}
A.Barrat, Phys.Rev. E \textbf{57}, (1998) 3629.

\bibitem{Parisi99}
G.Parisi, F.Ricci-Terzenghi and J.J.Ruiz-Lorenzo, 
Eur.Phys.J. B \textbf{11}, (1999) 317.

\bibitem{Godreche00}
C.Godreche and J.M.Luck, J.Phys. A \textbf{33}, (2000) 1151.

\bibitem{Lippiello00}
E.Lippiello and M.Zannetti, Phys.Rev. E \textbf{61}, (2000) 3369.

\bibitem{Bray89}
A.J.Bray, J.Phys. A \textbf{22}, (1989) L67.

\bibitem{Prados97}
A.Prados, J.J.Brey and B.Sanchez-Rey, Europhys.Lett. \textbf{40}, (1997) 13.

\bibitem{Corberi01}
A first partial account of our results has been published in
F.Corberi, E.Lippiello and M.Zannetti, Phys.Rev. E \textbf{63}, (2001) 061506.

\bibitem{Palmer82}
R.G.Palmer, Adv.Phys. \textbf{31}, (1982) 669.

\bibitem{Kurchan96}
J.Kurchan and L.Laloux, J.Phys. A \textbf{29}, (1996) 1929; C.M.Newman
and D.L.Stein, J.Stat.Phys. \textbf{94}, (1999) 709.

\bibitem{Bray94}
For a review see A.J.Bray, Adv.Phys. \textbf{43}, (1994) 357.

\bibitem{Furukawa89}
H.Furukawa, J.Stat.Soc.Jpn. \textbf{58}, (1989) 216; Phys.Rev. B \textbf{40},
(1989) 2341.

\bibitem{Ohta82}
T.Ohta, D.Jasnow and K.Kawasaki, Phys.Rev.Lett. \textbf{49}, (1982) 1223.

\bibitem{Kawasaki78}
K.Kawasaki, M.C.Yalabik and J.D.Gunton, Phys.Rev. A \textbf{17},
(1978) 455.

\bibitem{Nagai86}
T.Nagai and K.Kawasaki, Physica A \textbf{134}, (1986) 483;
A.D.Rutenberg and A.J.Bray, Phys.Rev. E \textbf{50}, (1994) 1900.

\bibitem{Glauber63}
R.J.Glauber, J.Math.Phys. \textbf{4}, (1963) 294.

\bibitem{Cugliandolo94}
L.F.~Cugliandolo, J.~Kurchan and G.~Parisi, J.Phys.I France 
\textbf{4}, (1994) 1641.

%
%
\end{thebibliography}
\end{document}